\documentclass[apj]{emulateapj}
\usepackage{graphicx}
\usepackage{natbib}
\usepackage{amsmath}

\begin{document}

\newcommand{\civ}{\ion{C}{4}}
\newcommand{\alphauv}{$\alpha_{\rm uv}$}
\newcommand{\aox}{$\alpha_{\rm ox}$}
\newcommand{\ax}{$\alpha_{\rm x}$}
\newcommand{\luvaox}{$L_{\rm uv}$--$\alpha_{\rm ox}$}
\newcommand{\lbol}{$L_{\rm bol}$}
\newcommand{\luv}{$L_{\rm uv}$}
\newcommand{\lx}{$L_{\rm x}$}
\newcommand{\llam}{$L_{\lambda}$}
\newcommand{\ltwofive}{$L_{\rm 2500\,\AA}$}
\newcommand{\lfiveone}{$L_{\rm 5100\,\AA}$}
\newcommand{\bctwofive}{BC$_{\rm 2500\,\AA}$}
\newcommand{\bcfiveone}{BC$_{\rm 5100\,\AA}$}
\newcommand{\onemum}{1\,$\mu$m}
\newcommand{\tenmum}{10\,$\mu$m}
\newcommand{\thirtymum}{30\,$\mu$m}
\newcommand{\twokev}{2\,keV}
\newcommand{\tenkev}{10\,keV}
\newcommand{\hundredev}{0.1\,keV}
\newcommand{\twohundredev}{0.2\,keV}

\title{Mean Spectral Energy Distributions and Bolometric Corrections for Luminous Quasars}

\author{
Coleman M.\ Krawczyk,\altaffilmark{1}
Gordon T.\ Richards,\altaffilmark{1}
Sajjan S.\ Mehta,\altaffilmark{1}
Michael S.\ Vogeley, \altaffilmark{1}
S.~C. Gallagher,\altaffilmark{2}
Karen M.\ Leighly,\altaffilmark{3}
Nicholas P.\ Ross,\altaffilmark{4}
and
Donald P.\ Schneider \altaffilmark{5,6}
}

\altaffiltext{1}{Department of Physics, Drexel University, 3141 Chestnut Street, Philadelphia, PA 19104, USA}
\altaffiltext{2}{Department of Physics \& Astronomy, The University of Western Ontario, London, ON N6A 3K7, Canada.}
\altaffiltext{3}{Homer L. Dodge Department of Physics and Astronomy, The University of Oklahoma, 440 W. Brooks St., Norman, OK 73019, USA}
\altaffiltext{4}{Lawrence Berkeley National Laboratory, 1 Cyclotron Road, Berkeley, CA 94720, USA}
\altaffiltext{5}{Department of Astronomy and Astrophysics, The Pennsylvania State University, 525 Davey Laboratory, University Park, PA 16802, USA}
\altaffiltext{6}{Institute for Gravitation and the Cosmos, The Pennsylvania State University, University Park, PA 16802, USA}

\begin{abstract}

We explore the mid-infrared (mid-IR) through ultraviolet (UV) spectral energy distributions (SEDs) of 119,652 luminous broad-lined quasars with $0.064<z<5.46$ using mid-IR data from {\em Spitzer} and {\em WISE}, near-infrared data from Two Micron All Sky Survey and UKIDSS, optical data from Sloan Digital Sky Survey, and UV data from Galaxy Evolution Explorer.  
The mean SED requires a bolometric correction (relative to 2500\,\AA) of \bctwofive$=2.75\pm0.40$ using the integrated light from \onemum--\twokev, and we further explore the range of bolometric corrections exhibited by individual objects. In addition, we investigate the dependence of the mean SED on various parameters, particularly the UV luminosity for quasars with $0.5\lesssim z\lesssim3$ and the properties of the UV emission lines for quasars with $z\gtrsim1.6$; the latter is a possible indicator of the strength of the accretion disk wind, which is expected to be SED dependent.
Luminosity-dependent mean SEDs show that, relative to the high-luminosity SED, low-luminosity SEDs exhibit a harder (bluer) far-UV spectral slope ($\alpha_{{\rm UV}}$), a redder optical continuum, and less hot dust.
Mean SEDs constructed instead as a function of UV emission line properties reveal changes that are consistent with known Principal Component Analysis (PCA) trends.
A potentially important contribution to the bolometric correction is the unseen extream-UV (EUV) continuum. 
Our work suggests that lower-luminosity quasars and/or quasars with disk-dominated broad emission lines may require an extra continuum component in the EUV that is not present (or much weaker) in high-luminosity quasars with strong accretion disk winds. As such, we consider four possible models and explore the resulting bolometric corrections.  
Understanding these various SED-dependent effects will be important for accurate determination of quasar accretion rates.

\end{abstract}

\keywords{catalogs --- infrared: galaxies --- methods: statistical --- quasars: general}

\section{Introduction}

Most massive (bulge-dominated) galaxies are believed to harbor a supermassive black hole at their centers \citep{Kormendy95}.  For the most part, these black holes are passive, but when the host galaxy's gas loses angular momentum it accretes onto the black hole, creating a luminous quasar \citep{Lynden-Bell69}.  This gas can be disrupted in a number of ways, including galaxy mergers \citep[e.g.][]{Kauffman00, Hopkins06} and secular fueling \citep[e.g.][and references therein]{Hopkins09}.
Most of the energy output is due to an accretion disk that forms about the black hole which either directly or indirectly produces radiation across nearly the entire electromagnetic spectrum (EM) \citep{Elvis94}.  
As a result of the quasar process, some of the energy is ``fed back'' into the galaxy, possibly disrupting the galaxy's gas supply \citep{Silk98,Fabian99} and quenching accretion of matter onto the black hole itself \citep[e.g.,][]{Matteo05, Djorgovski08}.  

Theoretical models of quasar feedback make predictions that are based on {\em bolometric luminosity} \citep[e.g., ][]{Hopkins06}, while what can be measured is usually a {\em monochromatic luminosity}, \llam.  The bolometric luminosity, \lbol, is given by the integrated area under the full spectral energy distribution (SED).  The ratio between \lbol\ and \llam\ defines a ``bolometric correction'', which can be applied generically if there is reason to believe that the quasar SED is well-known (if not well-measured for an individual object).  Currently the best quasar SEDs are based on only tens or hundreds of bright quasars \citep{Elvis94,Richards06}. However, over 100,000 luminous broad-line quasars have been spectroscopically confirmed \citep{Schneider10}.  As it is dangerous to assume that we can extrapolate the results from a few hundred of the brightest quasars to the whole population, it is important to expand our knowledge to cover the full quasar sample.

While \citet{Elvis94} and \citet{Richards06} provide the most complete SEDs in terms of number of objects and overall wavelength coverage, further understanding of the SED comes from a variety of other investigations from both spectroscopy and multi-wavelength imaging across the EM spectrum.  
For example, composite SEDs, from the radio to the X-ray, for 85 optically and radio selected bright quasars ($\log{(\nu L_{\nu})}\mid_{\lambda=3000 {\rm \AA}} \geq 44$) with $z<1.5$ were made by \citet{Shang11}, and the bolometric corrections for this sample were tabulated by \citet{Runnoe12}.  
\citet{Stern12} construct SEDs for over 3500 low-z ($z<0.2$) type 1 active galactic nucleus (AGN) with broad H$\alpha$ lines ranging from the near-infrared (near-IR) up to the X-ray with $\log{(\nu L_{\nu})}\mid_{\lambda=2500 {\rm \AA}} \geq 42$, while \citet{Assef10} present SEDs using over 5300 AGNs spanning the mid-infrared (mid-IR) up to the X-ray, with redshift going up to $\sim$ 5.6, from the AGN and Galaxy Evolution Survey \citep[AGES; ][]{Kochanek12}.
While in the mid-IR, \citet{Deo11} produced a composite spectrum using 25 luminous type 1 quasars at $z \sim 2$.

\citet{Vanden01} used 2200 Sloan Digital Sky Survey (SDSS) spectra in the redshift range of $0.04<z<4.79$ to construct a mean quasar spectrum covering wavelengths from 800\AA\ to 8555\AA\ in the rest frame.  From this they found that the mean UV continuum is roughly a power-law with $\alpha_{\nu}=-0.44$ ($f_{\nu}\propto\nu^{\alpha}$).    
To explore the far-UV (FUV) region of the quasar spectrum, \citet{Telfer02} used $\sim 330$ {\em Hubble Space Telescope} spectra of 184 quasars, with $z>0.33$, covering rest frame wavelengths from 500\AA\ to 1200\AA, and found an anti-correlation between the spectral index of the FUV ($\alpha_{{\rm FUV}}$) and the luminosity at 2500\AA.  This work was later supplemented at lower redshift and lower luminosity by \citet{Scott04}, who used 100 {\em FUSE} spectra covering the FUV.  By combining their data with that from \citet{Telfer02}, they found an anti-correlation that can be characterized by
\begin{equation}
 \alpha_{{\rm FUV}} = 21.02-0.49 \log \left( \frac{\lambda_{1100{\rm \AA}} L_{1100{\rm \AA}}}{\mbox{erg s}^{-1}}\right).
 \label{scott_eq}
\end{equation}


At shorter wavelengths, using 73 quasars, \citet{Avni82} found a dependency of the spectral index between the optical and the X-ray and a quasar's luminosity.  This relation was further studied by a number of authors, including \citet{Steffen06}, who used 333 quasars with $z<6$ and $\log{(\nu L_{\nu})}\mid_{\lambda=2500 {\rm \AA}} \geq 42$, to find the relationship between the UV and X-ray luminosities to be:
\begin{equation}
 \log{(L_{{\rm 2keV}})} = (0.721 \pm 0.011) \log{(L_{2500 {\rm \AA}})} + (4.531 \pm 0.688),
 \label{just_eq}
\end{equation}
while \citet{Just07} have extended this result to higher luminosities.
Recently, \citet{Lusso10} completed a similar study using 545 X-ray selected quasars and found a similar relationship.   In the X-ray regime, the mean SED appears to have $\alpha_{\nu}\sim-1$ \citep[e.g.,][]{George00} before cutting off at $\sim 500\,{\rm keV}$ \citep{Zdziarski95}.

To improve our understanding of the mean SED (and its range), we have constructed quasar SEDs, spanning from the mid-IR ($\sim$\thirtymum\ in the rest frame) to the FUV ($\sim$300\,\AA\ in the rest frame), for quasars cataloged by the SDSS \citep{York00}. The data used for this analysis are presented in Section \ref{data}.  Section \ref{corrections} describes all the corrections applied to the data. Section \ref{sed} gives an overview of our data analysis and construction of mean SEDs. Section \ref{bol} presents our findings for individual and mean bolometric corrections, and our discussion and conclusions are presented in Section \ref{EUV} and Section \ref{conclusions} respectively. Throughout this paper we use a $\Lambda$CDM cosmology with $H_0=71$ km s$^{-1}$ Mpc$^{-1}$, $\Omega_\Lambda = 0.734$, and $\Omega_m = 0.266$, consistent with the {\em Wilkinson Microwave Anisotropy Probe} 7 cosmology \citep{Jarosik11}.

\section{Data} \label{data}

\begin{figure*}
 \centering
 \begin{tabular}{cc}
 \includegraphics[width=3in]{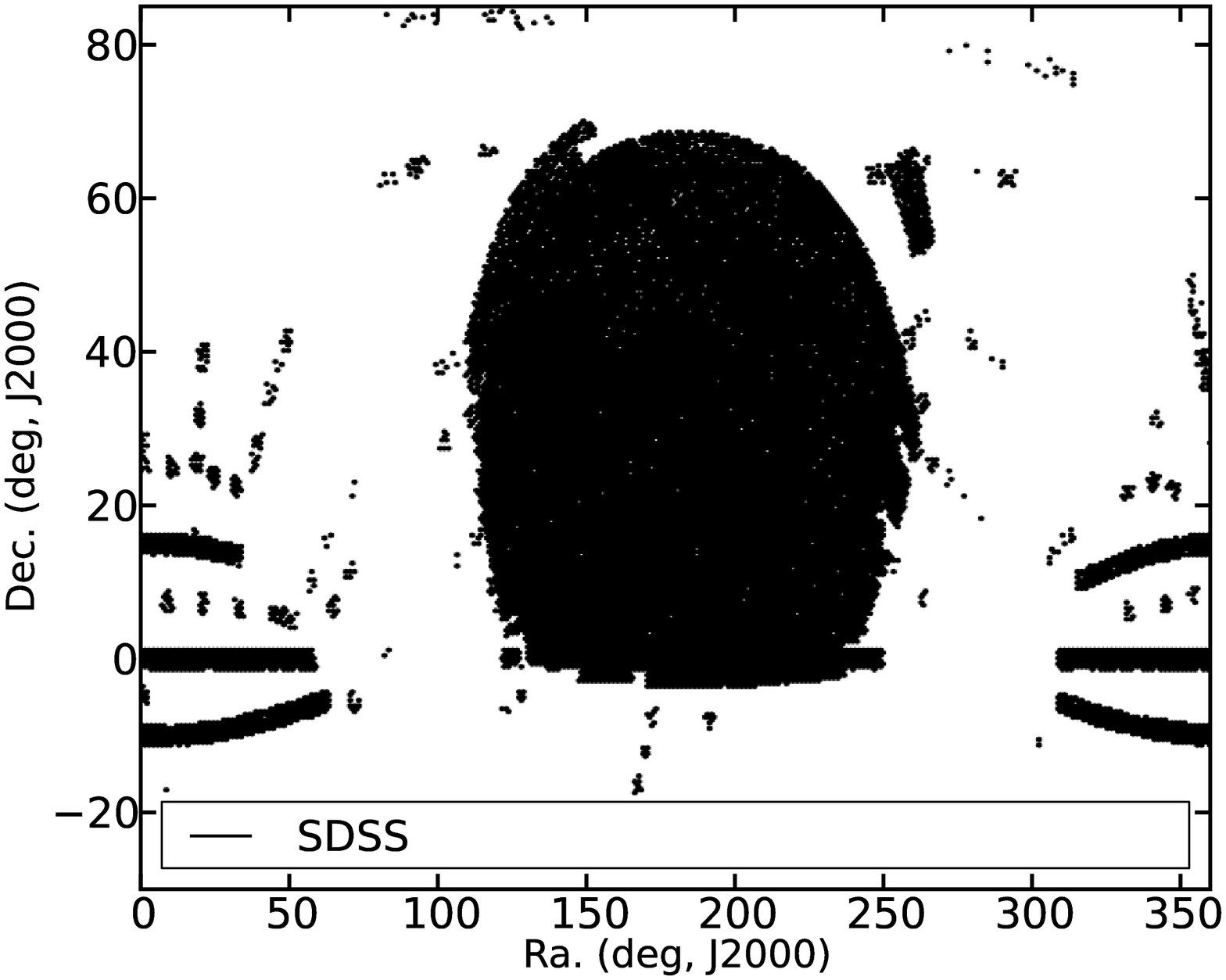} & \includegraphics[width=3in]{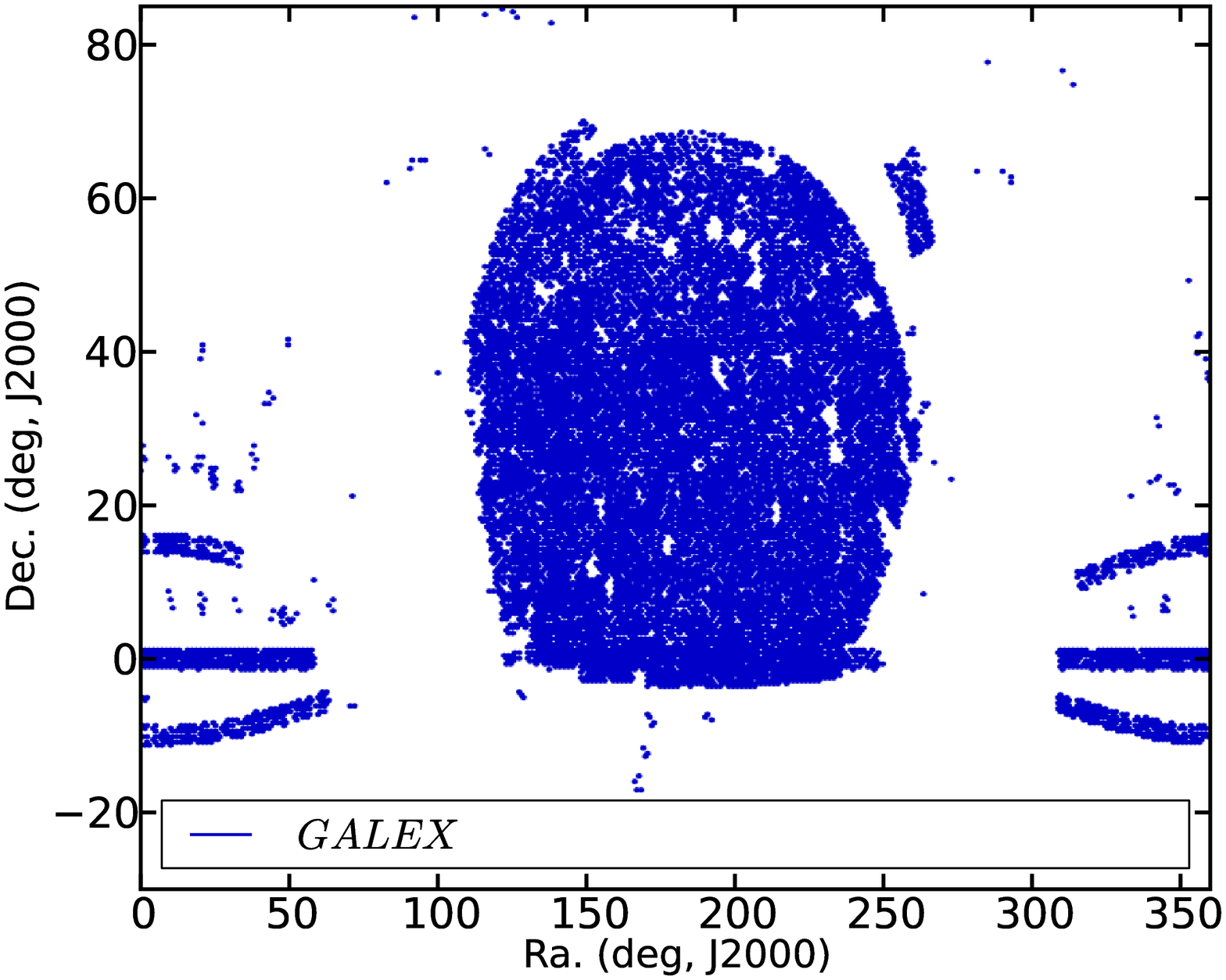} \\
 (a) SDSS quasar footprint & (b) {\em GALEX} quasar footprint \\
 \includegraphics[width=3in]{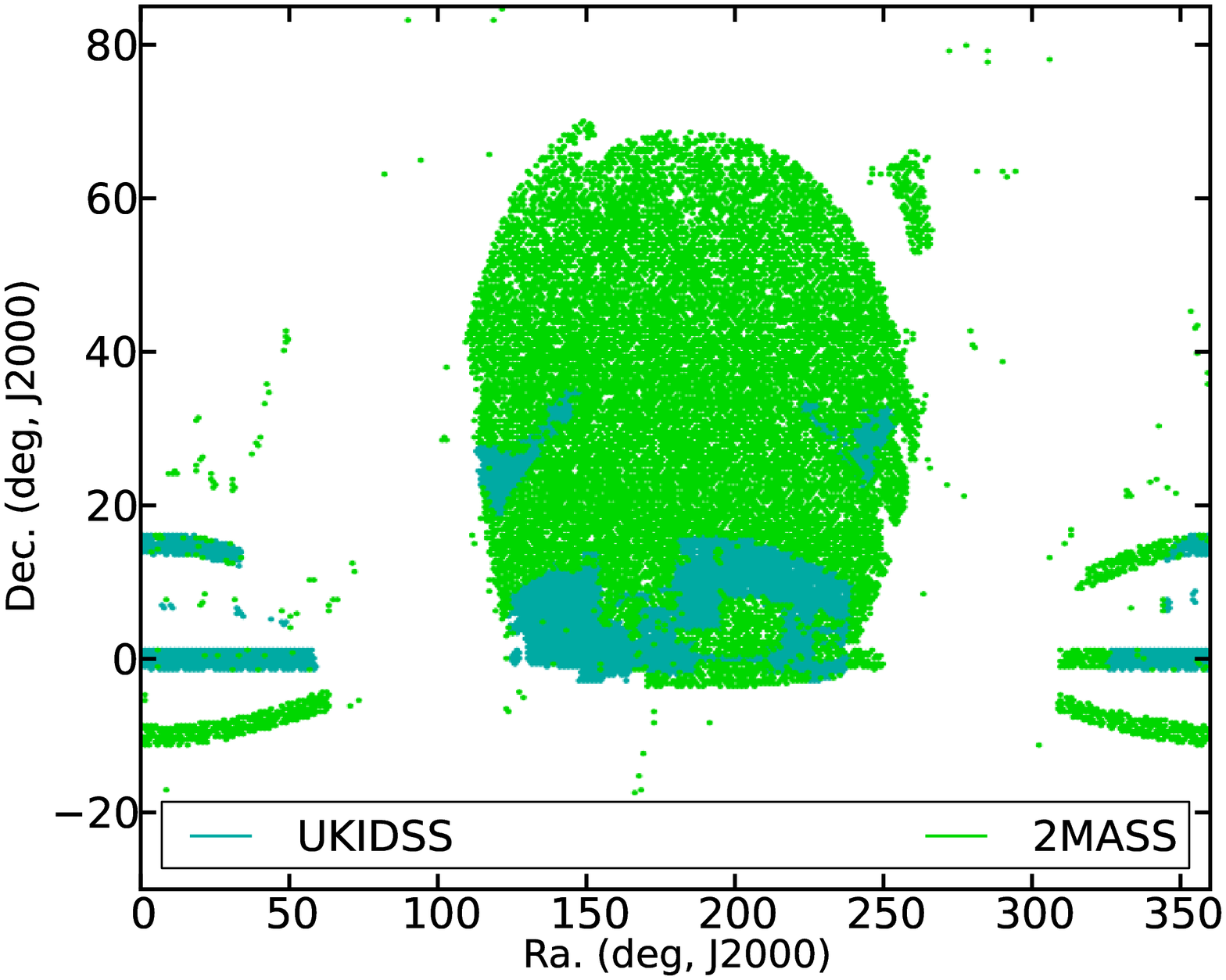} & \includegraphics[width=3in]{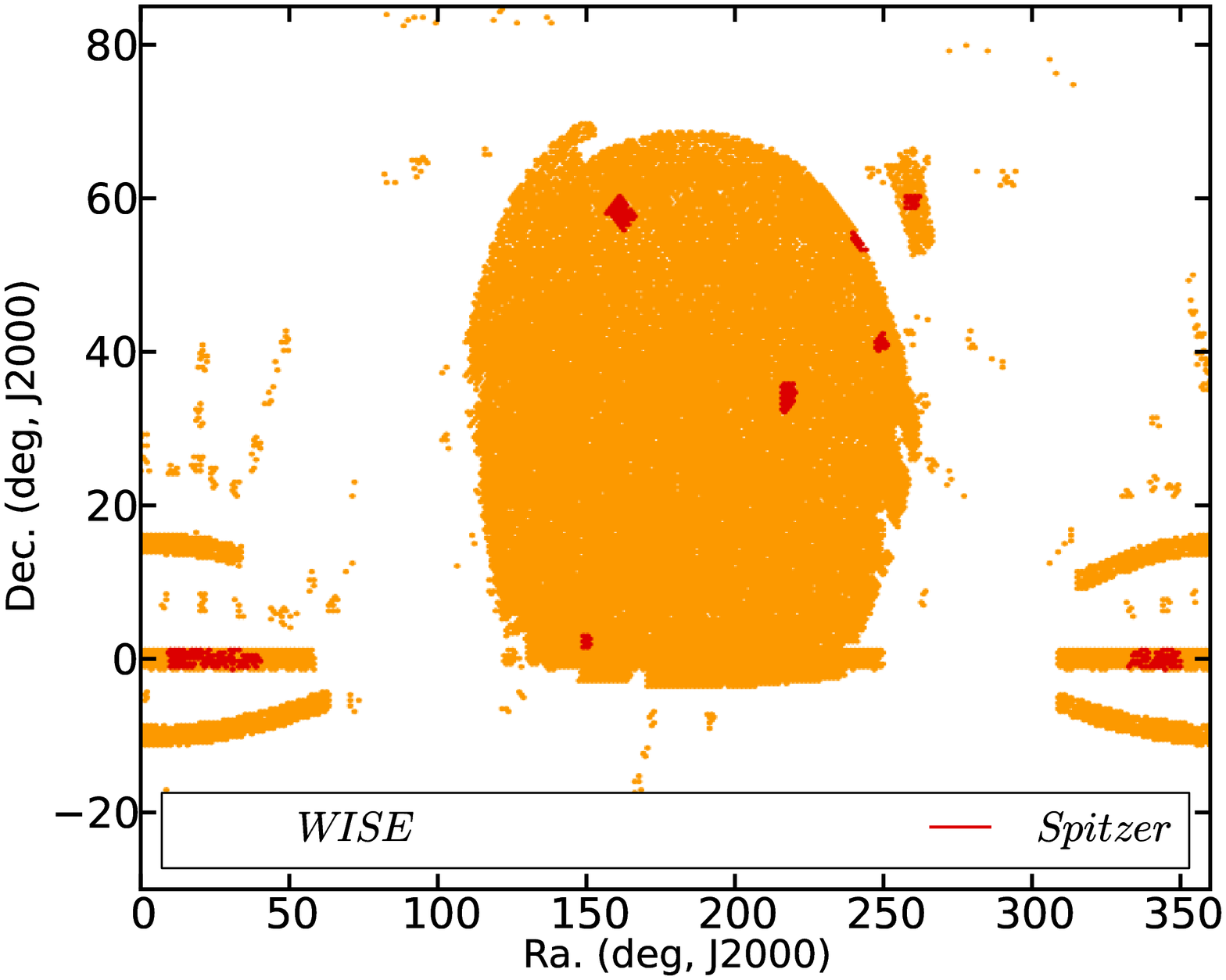} \\
 (c) Near-IR quasar footprint & (d) Mid-IR quasar footprint
 \end{tabular}
 \caption{Footprints of the optically selected quasars in the matched surveys.  Coverage of {\em GALEX} (blue), UKIDSS (teal), 2MASS (green), {\em WISE} (orange), and {\em Spitzer} (red) data are limited to the SDSS footprint (black) shown in the top-left panel.}
\label{foot}
\end{figure*}

\begin{figure}
\centering
\includegraphics[width=3.5in]{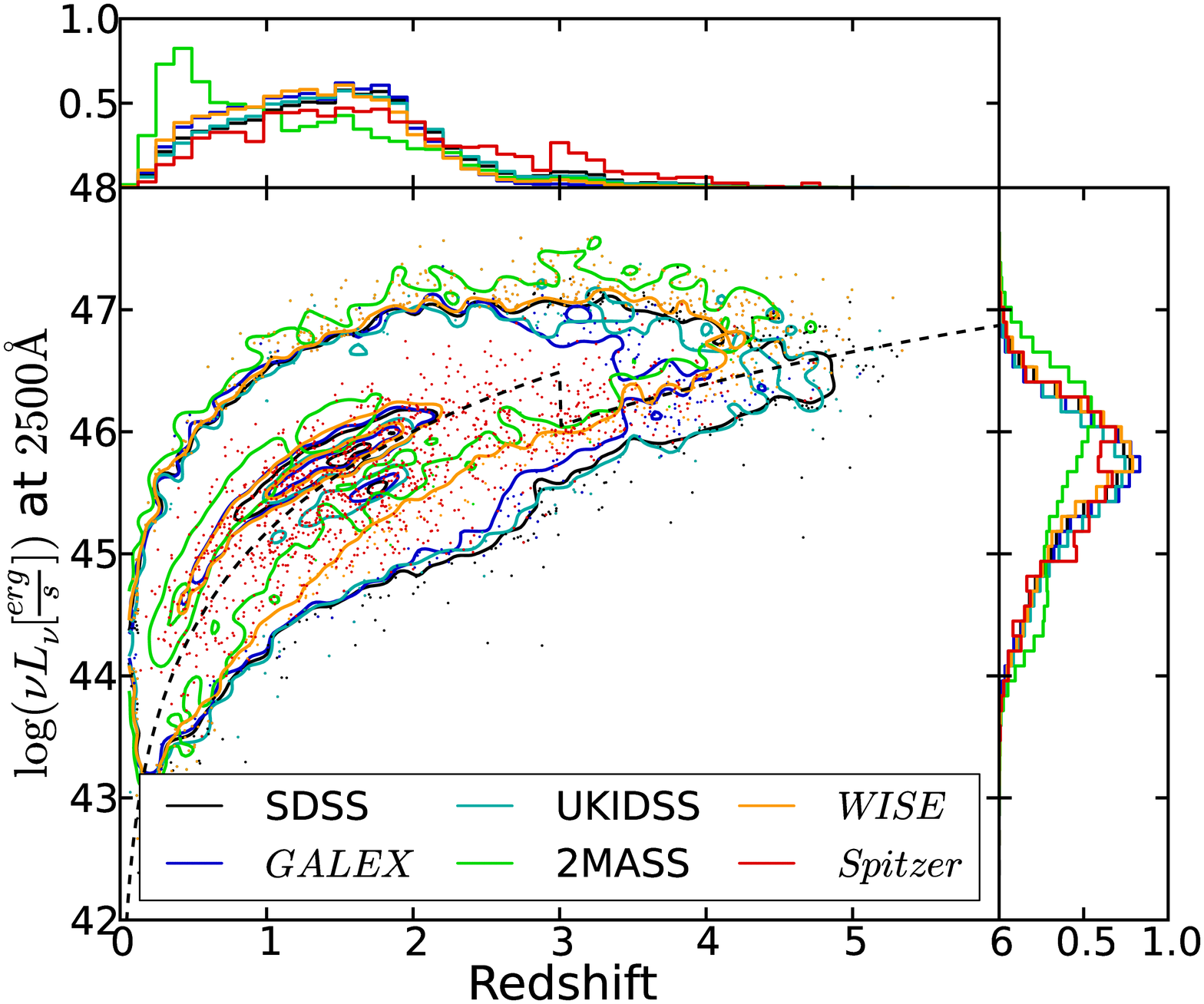}
\caption{Contour-scatter plot showing the 2500\,\AA\ luminosity vs. redshift for each quasar color coded by survey. The contours indicate the linear density of scatter points on the plot with level drawn at 0.005, 0.5, and 0.95 of the normalized distributions.  The histograms are normalized by the total number of quasars in each survey:  119,652 in SDSS, 42,043 in {\em GALEX}, 35,749 in UKIDSS, 23,088 in 2MASS, 85,358 in {\em WISE}, and 1196 in {\em Spitzer}.  The black dashed line represents the flux limit of the main SDSS survey before extending the sample to lower luminosities.}
\label{lum_v_z_survey}
\end{figure}

Our sample starts with the SDSS-DR7 quasar catalog by \citet{Schneider10}, containing 105,783 spectroscopically confirmed broad-lined quasars.  Each quasar has photometric data in the five SDSS optical bandpasses $ugriz$ \citep{Fukugita96}.  
Only quasars that have nonzero flux in all five SDSS bandpasses after correcting for Galactic extinction have been kept in our catalog, bringing the sample size to 103,895.
While the SDSS survey covers a large area of sky, it is limited to relatively bright quasars ($i<19.1$ for $z<3$ and $i<20.2$ for $z>3$).  As such, in addition to these quasars, we included 15,757 lower luminosity, optically-selected quasars taken from the Two Degree Field QSO Redshift survey \citep[2QZ; $b_J<20.85$ for $z<3$;][]{Croom04}, the Two Degree Field-SDSS LRG and QSO survey \citep[2SLAQ; $g<21.85$ for $z<3$;][]{Croom09}, and the AAT-UKIDSS-SDSS survey (AUS; $i<21.6$ for $0<z<5$; Croom et al., in preparation). 
In the rest frame, we have over 70,000 QSOs with $\log{(\nu [{\rm Hz}])}$ ranging from $\sim$13.7--15.4 ($\lambda\sim$1200\,\AA--6\,$\mu$m) and at least 10,000 QSOs covering the range from $\sim$13.3--15.6 ($\lambda\sim$750\,\AA--15\,$\mu$m).
All SDSS magnitudes have been corrected for Galactic extinction according to \citet{Schlegel98} with corrections to the extinction coefficients as given by \citet{Schlafly11}. 
 Table~\ref{data_table} presents the IR--UV photometry for the 119,652 quasars in our sample.  
 Signs of dust reddening are seen in 11,468 of these, specifically with a color excess $\Delta(g-i) > 0.3$ \citep{Richards03}. These have been excluded from our analysis, bringing our final sample size to 108,184. 
 Figure~\ref{foot} shows the sky coverage of our multi-wavelength data and Figure~\ref{lum_v_z_survey} shows the 2500\,\AA\ luminosity versus redshift distributions for each survey.
Note that, in the so-called SDSS ``Stripe 82'' region \citep{Annis11} along the Northern Equatorial Stripe, the SDSS imaging data reaches roughly 2 mag deeper than the main survey as a result of co-adding many epochs of data.  

\begin{deluxetable}{ll}
\tabletypesize{\scriptsize}
\tablecaption{Quasar Catalog Format \label{data_table}}
\tablehead{\colhead{Column} & \colhead{Description}}
\startdata
1   &   Previously published name\\
2   &   Right ascension in decimal degrees (J2000)\\
3   &   Declination in decimal degrees (J2000)\\
4   &   Redshift       \\
5   &    BEST SDSS {\em u} band PSF magnitude\\
6   &    Error in {\em u} magnitude\\
7   &    BEST SDSS {\em g} band PSF magnitude\\
8   &    Error in {\em g} magnitude  \\
9   &    BEST SDSS {\em r} band PSF magnitude\\
10  &    Error in {\em r} magnitude  \\
\enddata
\tablecomments{This table is published in its entirety in the electronic edition of the online journal. A portion is shown here
for guidance regarding its form and content.}
\end{deluxetable}

\subsection{Near-infrared}

Near-IR data in the {\em JHK} bandpasses are taken from the Two Micron All Sky Survey \citep[2MASS;][]{Skrutskie06}.  These include values that were matched to the 2MASS All-Sky and ``6$\times$'' point source catalogs using a matching radius of $2''$ with the 6$\times$ deeper catalog taking priority.  In addition to these data, we include 2$\sigma$ extractions (i.e., ``forced photometry'') at the positions of known SDSS quasars.  These objects have sufficiently accurate optical positions that it is possible to perform aperture photometry at their expected locations in the near-IR imaging, despite being non-detections in 2MASS.  These 2$\sigma$ ``detections'' were cataloged by \citet{Schneider10}; we include those objects with a signal-to-noise ratio (S/N) greater than 2.

To supplement the 2MASS data, we have matched our catalog to the UKIRT (United Kingdom Infrared Telescope) Infrared Deep Sky Survey \citep[UKIDSS; ][]{Lawrence07}. UKIDSS uses the $YJHK$ filter system \citep{Hewett06}; see also \citet{Peth11}.  
In order to generate our combined optical and near-IR data sets, we begin by source matching samples of SDSS data with the UKIDSS LAS catalog using the Cross-ID form located on the Web site of the WFCAM Science Archive (WSA)\footnote{\url{http://surveys.roe.ac.uk:8080/wsa/crossID\_form.jsp}}. In particular, we match against the UKIDSS DR5 LAS source table which contains the individual detections for a given object from each bandpass merged into a single entry.
We use a matching radius of $0\farcs7$ and the nearest neighbor pairing option, accepting only the nearest object with a detection in at least one band as an acceptable match.

\begin{figure*}
 \centering
 \begin{tabular}{cc}
 \includegraphics[width=3in]{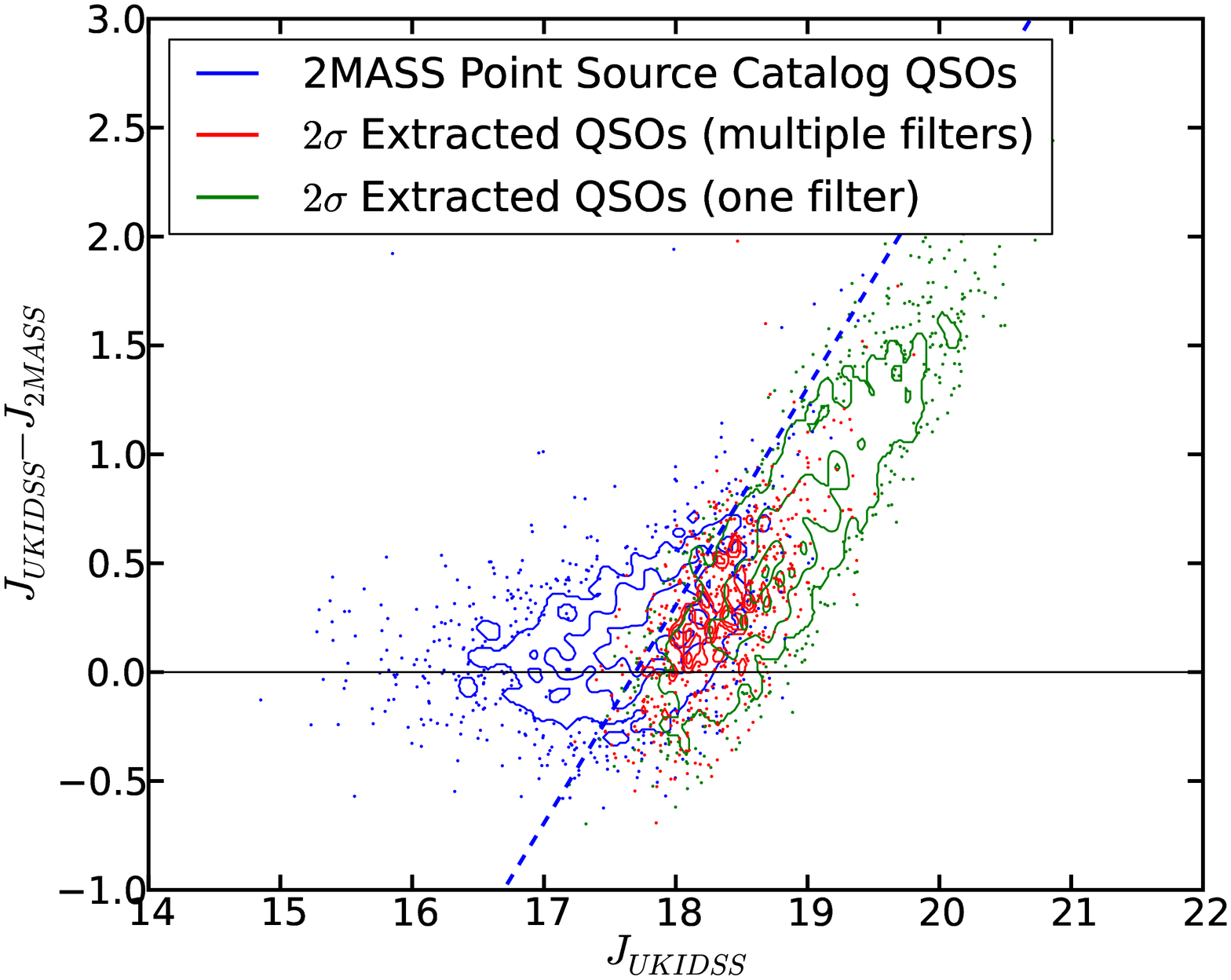} & \includegraphics[width=3in]{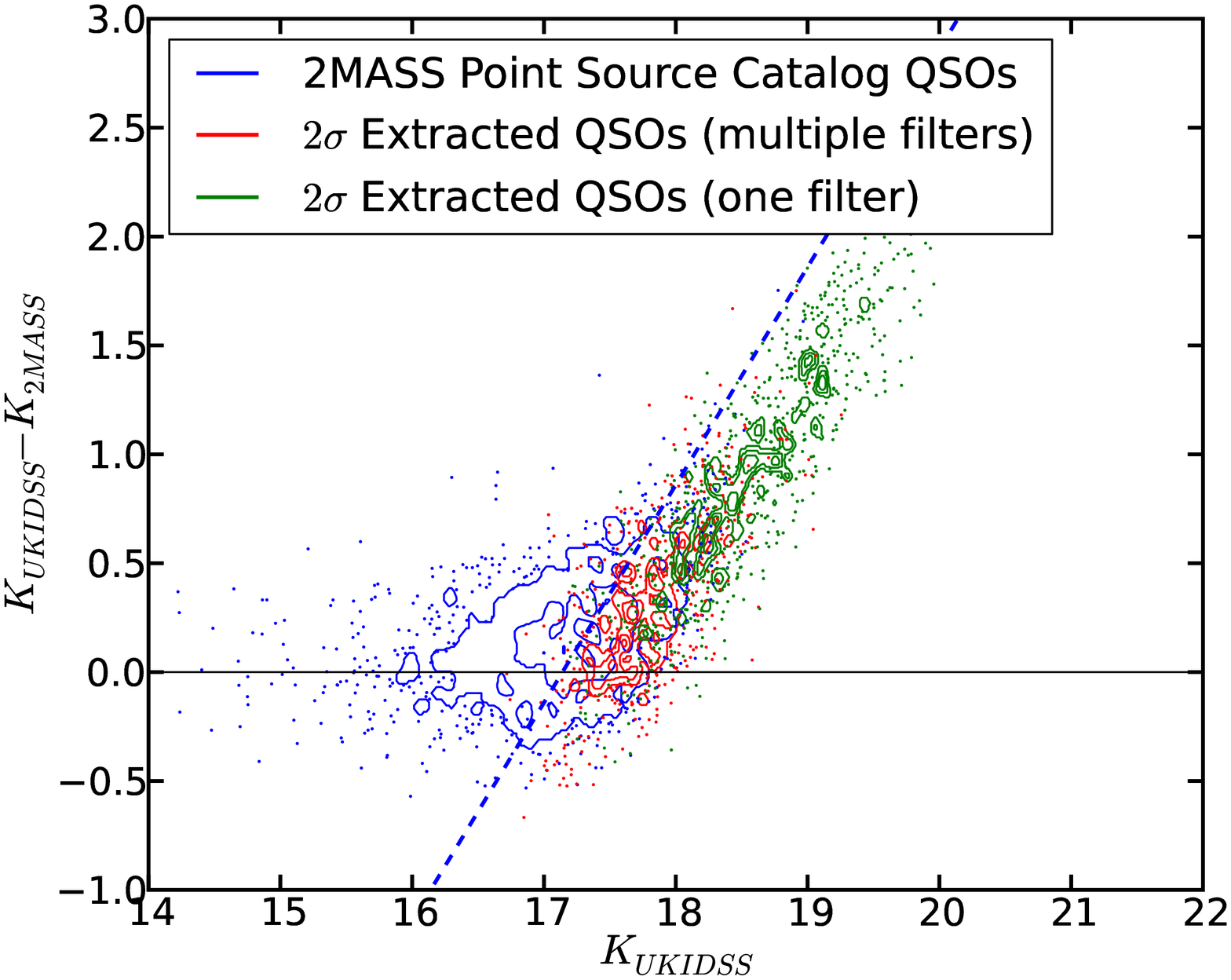}
 \end{tabular}
 
 \caption{Contour-scatter plot showing the difference between the 2MASS and UKIDSS magnitudes vs. the UKIDSS magnitudes for the $J$ and $K$ filters. The blue contours/dots show the catalog 2MASS values, green shows the 2MASS $2\sigma$ extraction values with a detection in only one band, and red shows the rest of the extracted values. The blue dashed line shows the 2MASS point source catalog's limiting magnitude.
 The two surveys contain 12,130 quasars in common.  Green points were excluded as the errors are quite large (see text).  Similar results were found for the $H$ filter.}
 \label{2mass_v_ukidss}
\end{figure*}

Since UKIDSS is deeper than 2MASS, when a quasar has data in both surveys, we only use the UKIDSS data. Between UKIDSS and 2MASS, 77,864 of our sample quasars have data in the near-IR: 35,749 from UKIDSS data and 41,434 from 2MASS, including both $5\sigma$ detections and forced photometry. Figure \ref{2mass_v_ukidss} shows the difference between the UKIDSS and 2MASS magnitudes versus the UKIDSS magnitudes for the 12,130 quasars the two surveys have in common.  We have split each figure into three distinct groups: quasars taken from the 2MASS $5\sigma$ catalog, quasars with $2\sigma$ 2MASS extraction values in more than one filter, and quasars with $2\sigma$ 2MASS extractions in only one filter.  As the quasars in the third group have much larger scatter (up to 2.5 mag) that can result in a misestimation of the quasar's luminosity by up to one order of magnitude,
we have removed all quasars that fall into this group.  This process leaves 23,088 objects in our 2MASS sample and brings our total number of QSOs with near-IR data to 58,837. 
In Table \ref{data_table} all 2MASS and UKIDSS fluxes have been converted to the AB magnitudes using \citet{Hewett06} conversions between AB and Vega magnitudes for the UKIDSS system.

\subsection{Mid-Infrared}

\begin{figure}
 \centering
 \includegraphics[width=3.5in]{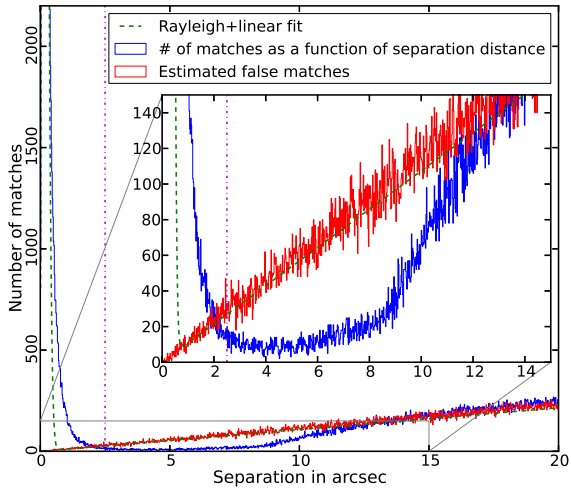}
 \caption{Angular separation between known quasars and nearby {\em WISE} sources (blue) and the known quasars with declinations shifted by 200$''$ (red).   The green curve shows the best fit expected distribution (Rayleigh distribution for small separations and linear growth for larger separations). The vertical magenta dot-dash line indicates the matching radius used.  As seen in the inset, the blue histogram does not follow the expected distribution; this is due, in part, to the large beam size of {\em WISE} (FWHM=$6''$).}
 \label{wise_sep}
\end{figure}

To extend our SEDs into the mid-IR, we included matches to the {\em WISE} final data release \citep{Wright10}.  The {\em WISE} bandpasses are generally referred to as $W1$ through $W4$ and have effective observed frame wavelengths of  3.36, 4.61, 11.82, and 22.13$\mu$m, respectively, for a typical quasar SED.  The matching was performed by taking all non-contaminated {\em WISE} point sources within $2''$ of an SDSS quasar as a match. This matching radius maximizes the number of true objects matched while also minimizing the number of false matches. Figure \ref{wise_sep} shows the number of {\em WISE} matches found as a function of separation distance from SDSS quasars. To ensure only the best matches were included, we also required that all matched quasars have S/N $\geq 10$ in both $W1$ and $W2$. In total, we have 85,358 matches to {\em WISE}; 
we estimate the false matching rate to be 1\% based on the number of matches obtained after shifting all {\em WISE} data by $200''$ in declination (red line in Figure \ref{wise_sep}). 
Since the {\em WISE} photometry was calibrated against blue stars it tends to overestimate the flux of red sources by about 10\% in the $W4$ band \citep{Wright10}, this correction has been applied to all quasars in our sample.

When available, {\em Spitzer} IRAC data were also included.  We specifically included data from the Extragalactic First Look Survey \citep[XFLS;][]{Lacy05}; Spitzer  Deep-Wide Field Survey \citep[SDWFS;][]{Ashby09}; SWIRE DR2 \citep{Lonsdale03}, including the ELAIS-N1, -N2, and Lockman Hole fields; S-COSMOS \citep{Sanders07}, and our own extraction of high redshift QSOs in stripe 82 (S82HIZ, program number 60139).  The fluxes for known SDSS quasars from the Stripe 82 program are reported here for the first time in Table \ref{data_table} ({\em Spitzer} source labeled as S82HIZ).  The IRAC bandpasses (ch1--ch4) have effective wavelengths of 3.52, 4.46, 5.67, and 7.7 $\mu$m for a typical quasar SED.  There are a total of 1196 matches to {\em Spitzer}.  For a small number of quasars ($\sim200$) both {\em Spitzer} and {\em WISE} data are available; in these cases, we use both data sets.  In Table \ref{data_table} all {\em Spitzer} and {\em WISE} fluxes are reported in AB magnitudes.

\subsection{Ultraviolet}

To extend the SEDs into the UV, we use {\em Galaxy Evolution Explorer} \citep[{\em GALEX}; ][]{Martin05} data when available.  The effective wavelengths for the near-UV (NUV) and the FUV bandpasses are 2267\AA\ and 1516\AA, respectively. Most of our matches are taken from \citet{Budavari09} who matched {\em GALEX}-GR6 to SDSS-DR7.  For this study, we have taken only the most secure matches; that is, only one SDSS quasar matched to only one {\em GALEX} point source.  This algorithm gives 14,302 matches and leaves out 53,782 quasars that have multiple SDSS sources matching to the same {\em GALEX} source.  In addition to these data, we include forced point-spread function (PSF) photometry \citep{Bovy11,Bovy12} on the {\em GALEX} images in the positions of the SDSS point sources, which adds 61,490 detections.  As with 2MASS, we only kept the extracted data with S/N $\geq 2$, and as with the \citet{Budavari09} matches, we limit the sample to those with no source confusion; this reduces the number of extracted quasars to 27,744.  Combining both sets, the total number of matches is 42,046.  All {\em GALEX} photometry has been corrected for Galactic extinction, assuming $A_{{\rm NUV}}= 8.741\times E(B-V)$ and $A_{{\rm FUV}}= 8.24\times E(B-V)-0.67\times E(B-V)^2$ \citep{Wyder07}.  The $E(B-V)$ values are taken from the \citet{Schneider10} catalog.  All the matched data are given in Table~\ref{data_table}.

\subsection{X-Ray} \label{X-ray}

To construct full SEDs, we must extend our data into the X-ray regime.  Due to comparatively limited sky coverage of sensitive X-ray observations, the number of our quasars with X-ray data is quite small compared to the size of the samples in the optical and IR.  For example, there are only 277 matches in the ChaMP data set \citep{Green09}.
So instead we take advantage of the careful work on mean X-ray properties as a function of UV luminosity as compiled by \citet{Steffen06}.  Specifically, we determine the X-ray flux of each quasar using the $L_{\rm UV}$--$L_{\rm X}$ relation that parameterizes the correlation between the 2500\AA\ and \twokev\ luminosity of quasars. We find the 2500\AA\ luminosity by extrapolation from the closest filter using an $\alpha_{\rm opt}=-0.44$ \citep{Vanden01}.  The \twokev\ luminosity is estimated using Equation (\ref{just_eq}) and our 2500\AA\ luminosities.  An X-ray energy spectral index $\alpha_x=-1$ (photon index $\Gamma=-2$) was assumed between \twohundredev\ and \tenkev\ \citep[e.g. ][]{George00}.  In this way, we can estimate the X-ray part of the SED for all of our sources rather than using just the small fraction of objects with robust X-ray detections. This process ignores the correlations between \aox\ (and \alphauv) and \ax\ as discussed in \citet{Kruczek11}; however, these trends are small as compared to the overall trend with luminosity.  

As a check we have compared our X-ray extrapolations to ChaMP data from \citet[][see Section \ref{sed}]{Green09}. In all cases we found the X-ray data to fall within the 1.5$\sigma$ limits of the \aox\ extrapolation. 

\section{Corrections} \label{corrections}

When studying quasar SEDs, we are interested in the {\em true} continuum level of the radiated light.  The continuum may be contaminated by, for example, spectral emission lines, absorption by intergalactic hydrogen clouds, host galaxy contamination, and beaming effects.  Here we address each of these features.  For the hydrogen absorption and broad emission lines, we determine a magnitude correction by folding a model through the filter curves as discussed below.  For the host galaxy correction, we subtract a model template.  We do not consider beaming specifically, but refer the reader to \citet{Runnoe12} for a discussion of how our results would change under the assumption of non-isotropic emission; see also \citet{Nemmen10}.

\subsection{Lyman Forest and Limit}
\label{sec:hydrogen}

\begin{figure}
 \centering
 \includegraphics[width=3.5in]{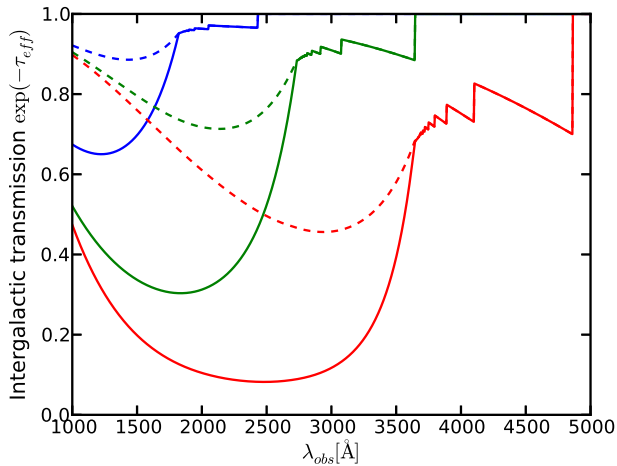}
 \caption{\citet{Meiksin06} model for Lyman series extinction as a function of observed wavelength for $z=1$ (blue), $z=2$(green), and $z=3$ (red). The solid lines show the extinction assuming a LLS (at the corresponding redshifts) along the line of sight, and the dashed lines show the extinction without a LLS. The jagged edges are a result of absorption from the Lyman series and the drop-off is due to absorption from the IGM and any hydrogen clouds along the line of sight. The transmission rises with decreasing wavelength for each feature because the shorter wavelengths sample lower redshifts and the universe is more ionized at lower redshifts.}
 \label{meiksin}
\end{figure}

A particular challenge is to determine the SED in the extreme-UV (EUV) ($\lambda_{\rm rest}<1216{\rm \AA}$) part of the spectrum where intergalactic hydrogen causes significant attenuation of the quasar signal \citep[Ly$\alpha$ forest;][]{Lynds71}.  To account for this attenuation, the redshift-dependent effective optical depth, $\tau_{{\rm eff}}(z,\lambda)$, of \citet{Meiksin06} was used.  
This optical depth models the average attenuation of a source assuming Poisson-distributed intergalactic hydrogen along the line of sight, out to a given redshift.
This optical depth is split into three parts: the contribution due to resonant scattering by Lyman transitions, systems with optically thin Lyman edges, and Lyman Limit Systems (LLS). Figure~\ref{meiksin} shows this model for redshifts of 1, 2, and 3 for both LLS and non-LLS.  

Because our investigation uses photometry and not spectra, we do not know whether a LLS is present.  Furthermore, we only have spectral coverage of the LLS region for a fraction of our sources.  Thus, to be as conservative as possible, we assume that there is {\em no} LLS present; that way we only make the {\em minimum} correction needed for each SED.  However, it is important to note that the SDSS quasar selection algorithm has been shown to preferentially select quasars possessing a LLS in the range $3 \lesssim z \lesssim 3.5$ \citep{Worseck11}.

For the continuum ($F_{\nu}$) we have assumed a power-law of the form $\nu^{-0.44}$ \citep{Vanden01}, consistent with the results of \citet{Scott04}.  Continuum weighting is necessary since we are dealing with broadband photometry and not spectroscopy.   With high resolution spectroscopic data the correction is exact as the \citet{Meiksin06} corrections themselves are independent of the SED.  However, for broadband photometry, where the bandpass can overlap features in the $\tau$ distribution, the convolution of the SED with the filter response changes the effective wavelength of the bandpass.

To apply the Lyman forest correction, we convolve a continuum with the \citet{Meiksin06} optical depth and each filter over $0<z<6$ in $\Delta z$ steps of 0.01 and use linear interpolation to precisely match redshifts. Following \citet{Meiksin06}, the correction is given in magnitudes as a transmission weighted average of $\tau$:
\begin{equation}
K_{\rm IGM} =  -2.5\times \log{\left( \frac{\int \lambda S_{\lambda} F_{\lambda} e^{-\tau_{{\rm eff}}} d\lambda}{\int \lambda S_{\lambda} F_{\lambda} d\lambda} \right)},
\end{equation} where $S_{\lambda}$ is a transmission filter curve, $F_{\lambda}$ is the continuum, and $\tau_{{\rm eff}}$ is the redshift-dependent effective optical depth. This value is calculated independently for each filter and makes $m(\tau=0) = m(\tau)-K_{\rm IGM}$ such that all quasars are made brighter in the Lyman absorption regions as a result of the correction.  

\subsection{Emission Lines}
\label{sec:emission}

\begin{figure}
 \centering
 \includegraphics[width=3.5in]{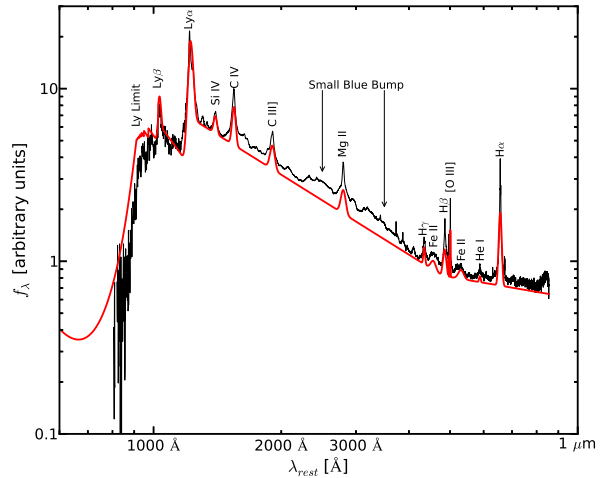}
 \caption{\citet{Vanden01} composite spectra (black) with our model spectrum with a LLS (red) at a redshift of 3.75.  Since the SDSS spectra cover 3800--9200\,\AA, the contributors to the points at $\lambda_{{\rm rest}}<1216$\,\AA\ are high-$z$ quasars and are consistent with the model used. Since we model the main emission line features, on average, they should not appear in the SEDs.  However, unmodeled residuals in \ion{Fe}{2} and the Balmer continuum (small blue bump) can clearly be seen in the SEDs on either side of the \ion{Mg}{2} emission line at 2800\,\AA. We use a double power law continuum in this figure since the composite spectra do not correct for host galaxy contributions on the red end.  Model spectra used to correct our SEDs are placed on a single power law as shown in Figure~\ref{mock}.}
 \label{van}
\end{figure}

The presence of emission lines in a photometric bandpass affects quasar magnitudes as illustrated by \citet[][e.g., see their Figures 8 and 17]{Richards06qlf}.  Thus, to make a fair comparison between sources at all redshifts, we endeavor to remove first-order effects of the emission line contributions to the measured magnitudes.  Our emission-line template takes the position, width, and equivalent width (EW) of the 13 strongest spectral lines (as labeled in Figure~\ref{van}) from \citet{Vanden01}.  We make no attempt to correct for the small blue bump (i.e. 
Balmer continuum or \ion{Fe}{2} emission), but recognize that those features can have a significant impact; indeed their residuals can be seen in our mean SEDs.  Figure~\ref{van} shows the mock spectrum used in our emission-line corrections; it includes a double power-law continuum ($\alpha_{\nu}=-0.46$ for $\lambda_{{\rm rest}} \leq 4600\,{\rm \AA}$ and $\alpha_{\nu}=-1.58$ for $\lambda_{{\rm rest}}>4600\,{\rm \AA}$) and 13 emission lines.  

The correction is computed as
\begin{equation}  
K_{\rm em} = -2.5\times \log{\left( \frac{\int \lambda S_{\lambda} F_{\lambda}(c\&l) d\lambda}{\int \lambda S_{\lambda} F_{\lambda}(c) d\lambda}\right)},
\end{equation}
where $S_{\lambda}$ is a transmission filter curve, $F_{\lambda}(c)$ is the continuum, and $F_{\lambda}(c\&l)$ is the continuum with spectral lines.
This value is calculated independently for each filter and makes $m(c) = m(c\&l)-K_{\rm em}$, such that all quasars are made fainter by the removal of the emission line contribution.

We further use this template to illustrate the correction for hydrogen absorption (assuming the quasar is at $z=3.75$ and there is a LLS).  Overall, we find reasonable agreement (by eye) with the mean composite spectrum from \citet{Vanden01} in terms of the continuum, broad emission lines, and intergalactic medium (IGM) attenuation.

\begin{figure}
 \centering
 \includegraphics[width=3.5in]{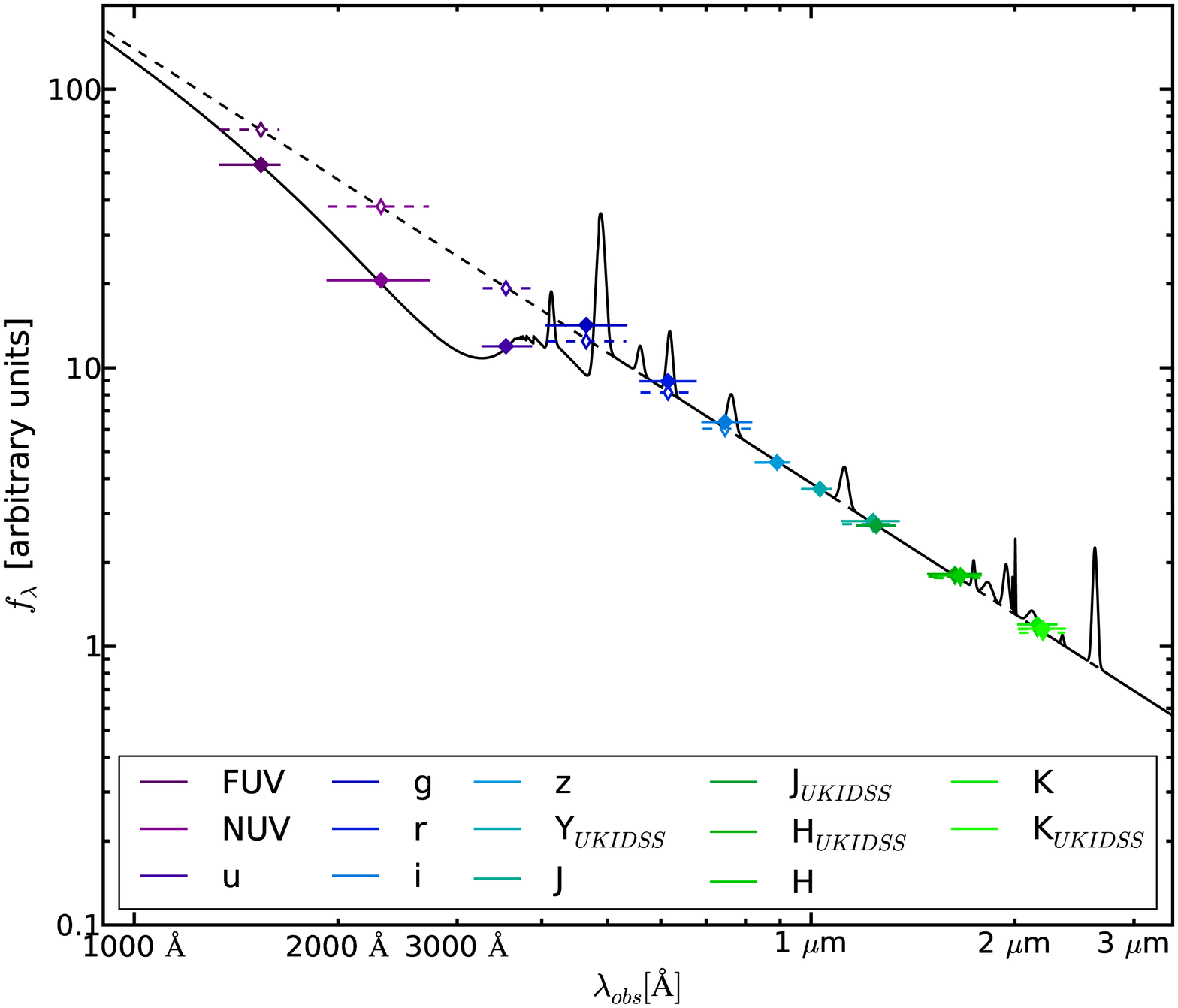}
 \caption{Observed frame mock spectrum (solid line) and $\alpha_{\nu}=-0.44$ continuum (dotted line) for a $z=3$ quasar.  The closed diamonds bifurcated by solid lines indicate the ``measured'' instrumental magnitudes, while open diamonds bifurcated by dashed lines indicated the corrected values (assuming an underlying power-law SED).  The width of the lines indicate the width of the bandpasses.   Here we have corrected for IGM attenuation (Section~\ref{sec:hydrogen}) and emission line flux (Section~\ref{sec:emission}).
}
 \label{mock}
\end{figure}

In Figure~\ref{mock}, we illustrate the effect of these emission-line and IGM corrections using a mock spectrum of a $z=3$ quasar (without a LLS).  Colored points indicate the observed instrumental magnitudes and the intrinsic continuum magnitudes after applying the emission line and IGM $K$-corrections.  

\subsection{Host Galaxy}
\label{sec:host}

It is also necessary to correct the data for host galaxy contamination.  Since we lack the data to measure this directly, we must estimate the host galaxy contribution.  To do this, we combine two different models: the relation outlined in \citet{Shen11} for the higher luminosity quasars, and the relationship from \citet{Richards06} for the lower luminosity quasars.  The \citet{Shen11} relationship is
\small
\begin{equation}
 \frac{L_{5100{\rm \AA,host}}}{L_{5100{\rm \AA,QSO}}}=0.8052-1.5502x+0.9121x^2-0.1577x^3 
 \end{equation}
 \normalsize
 where
 \small
 \begin{equation}
 x \equiv \log{ \left( \frac{\lambda_{5100{\rm \AA}} L_{5100{\rm \AA},{\rm total}}}{\mbox{erg s}^{-1}} \right) } - 44 \nonumber
 \end{equation}
 \normalsize
 and
 \small
 \begin{equation}
 0 \leq x < 1.053, \nonumber
\end{equation}
\normalsize
which sets the relative scaling at 5100\AA. To extend this expression to all wavelengths, we use the elliptical galaxy template of \citet{Fioc97} scaled to 5100\AA.  When $x \geq 1.053$, the quasar completely outshines the host galaxy and no correction needs to be applied.  We note that this relationship was found using quasars that have $\log{(\lambda_{5100{\rm \AA}} L_{5100{\rm \AA},{\rm total}})} \geq 44$, and cannot, in general, be extrapolated to lower luminosities.

We found that, for lower luminosity quasars, the \citet{Shen11} equation overestimates the host galaxy; we instead use the relationship from \citet{Richards06} which, was adapted from the relationship used by \citet{Vanden06}:
\footnotesize
\begin{eqnarray}
\log{(L_{6156{\rm \AA},host})} &=& 0.87 \log{(L_{6156{\rm \AA},AGN})} \nonumber \\
	&& + 2.887 - \log{(L_{{\rm Bol}}/L_{{\rm Edd}})} \\
\mbox{where } \log{(L_{6156{\rm \AA},AGN})} &=& \log{\left( \frac{L_{6156{\rm \AA},total}}{\mbox{erg s}^{-1} \mbox{Hz}^{-1}} - \frac{L_{6156{\rm \AA},host}}{\mbox{erg s}^{-1} \mbox{Hz}^{-1}} \right)}. \nonumber
\end{eqnarray}
\normalsize
This system of equations is then solved numerically for the host galaxy luminosity.  As in \citet{Richards06} we take $L_{{\rm Bol}}/L_{{\rm Edd}}$ to be unity, since this provides the minimum correction needed.  This sets the relative scaling at 6156\AA\, and, as before, we use the elliptical galaxy template of \citet{Fioc97} to extend this to all wavelengths.  To ensure a smooth transition between these two methods, we have chosen the crossover luminosity to be the point at which both methods agree, $\log{(\lambda_{5100{\rm \AA}} L_{5100{\rm \AA},{\rm total}})}=44.75$.
As host galaxy subtraction can have a large impact on the SED near 1\,$\mu$m and it is not a well-established procedure, we note that
other methods include \citet{Croom02}, and \citet{Maddox06}.

It was brought to our attention that the host galaxy model used in \citet{Richards06} was incorrect (K.\ Leighly, private communication 2012).  The model was converted from $F_{\lambda}$ to $F_{\nu}$, but was not converted to $\nu F_{\nu}$ before subtracting it from the SEDs.  This does not effect the SED at 5100\AA, where the SEDs are normalized, but it is systematically wrong on either side of this wavelength.   Herein we correct that error.

\subsection{Gap Repair}

Since all the quasars in our sample have been selected from SDSS photometric data, they all have $ugriz$ measurements, but are not guaranteed to have measurements in the other bandpasses that we utilize. To address this issue we will ``gap repair'' all missing data in a way similar to \citet{Richards06}.  Specifically, we replace missing values with those determined by normalizing an interpolation/extrapolation of the continuum in the next nearest bandpass for which we have data.   Sometimes this will be a previously constructed mean SED; other times it will be a functional form.
To estimate errors for the gap filled points, we have fit up to third-order polynomials to plots of the magnitude errors, $\log(\sigma_m)$, versus magnitudes, $m$, for each of the filters.  These functions were then used to estimate $\sigma_m$ for the gap filled value for $m$.

This procedure works well where we can interpolate between filters (e.g., at wavelengths longward of the Lyman limit).  Beyond the Lyman limit, we are no longer interpolating, but extrapolating, which is a trickier process.  Using limiting magnitudes does not solve the problem as redshift effects mean that even if all quasars had coverage from all of the filters, the same rest-frame wavelengths are not observed for both high- and  low-redshift quasars.

On the long wavelength end, we always extrapolate to longer wavelengths using the mean SED from \citet{Richards06}, normalized to the nearest measured bandpass.  On the short wavelength end, the gap repair process depends on whether we are making a mean SED or attempting to reconstruct the SED for an individual quasar (e.g., to determine its bolometric luminosity).  See below for details for specific cases.

Unless otherwise stated, the X-ray part of the SED is determined using Equation~(\ref{just_eq}), the \luvaox\ relation from \citet{Steffen06}, and linearly interpolating between the X-ray and the bluest data point after gap repair. The errors in this region are determined from the uncertainties of the \citet{Steffen06}  relation.

\section{Mean SEDs} \label{sed}

\subsection{Overall Mean SED}
To determine the mean quasar SED, we converted all the flux densities for each quasar to luminosity densities and shifted each broadband observation to the rest frame.  The data were then placed onto a grid with points separated by 0.02 in $\log{(\nu)}$; 
kriging was then used to align the broadband luminosities to our grid points.  Kriging is a nonparametric interpolation method that predicts values and errors for regions between observed data points.  \citet{Rybicki92} were among the first to present this technique in the astronomical literature (under the name Wiener filtering).  Since then it has been mainly used to estimate light curves \citep[e.g.,][]{Kozlowski10}.

Kriging 
estimates the correlation between data points as a function of separation, then uses this correlation to interpolate between the data points.  
We used the \texttt{R} package \texttt{gstat} to perform variance-weighted kriging using an exponential variogram model.
Once the data are rebinned in the rest-frame, the arithmetic mean and standard deviation are taken at each grid point.  
Because kriging estimates variance values at each of the new grid points, we are able to estimate errors on our bolometric luminosities and the derived bolometric corrections. 

\begin{figure}
 \centering
 \includegraphics[width=3.5in]{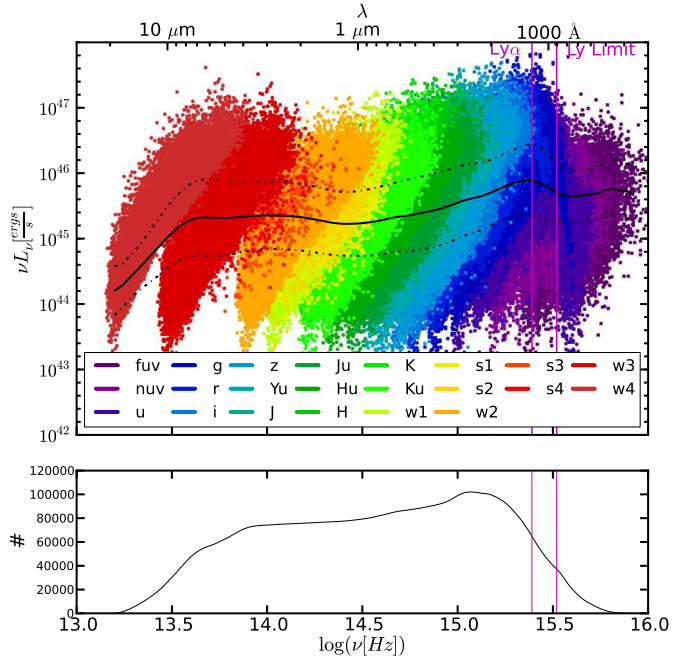}
 \caption{
 Top: mean uncorrected SED (solid line), data points for all quasars, and 1$\sigma$ level (dash-dotted line) for the raw sample (no corrections applied, see Section~\ref{corrections}). The steep drop-off on the red end is caused by the lack of high luminosity, low redshift quasars in our sample; this is caused by both quasar evolution and the fact that at lower redshifts we sample a smaller volume of the universe. The white wedges between the filters are caused by the limiting magnitudes for each filter. The drop-off just shortward the Ly$\alpha$ line is due to IGM attenuation, but the flattening shortward of the Lyman limit is due to the lack of low-luminosity, high-redshift quasars, caused by the detection limits of the surveys.
 Bottom: the number of SEDs averaged at each frequency. The two vertical magenta lines show the positions of the Ly$\alpha$ line at 1216\,\AA\ and the Lyman limit at 912\,\AA.
 }
 \label{mean_no_cor}
\end{figure}

Figure \ref{mean_no_cor} shows the ``raw'' data points (without the corrections described in Section~\ref{corrections}) and mean for all the quasars in our sample.  On the red end, the SED drops due to a lack of data and on the blue end, shortward of the Ly$\alpha$ line, there is a clear drop due to intergalactic extinction.  The flattening in the mean shortward of the Lyman limit is due to redshift effects; low-redshift sources have no rest-frame measurements at this frequency, biasing the data toward higher luminosity.  

\begin{figure}
 \centering
 \includegraphics[width=3.5in]{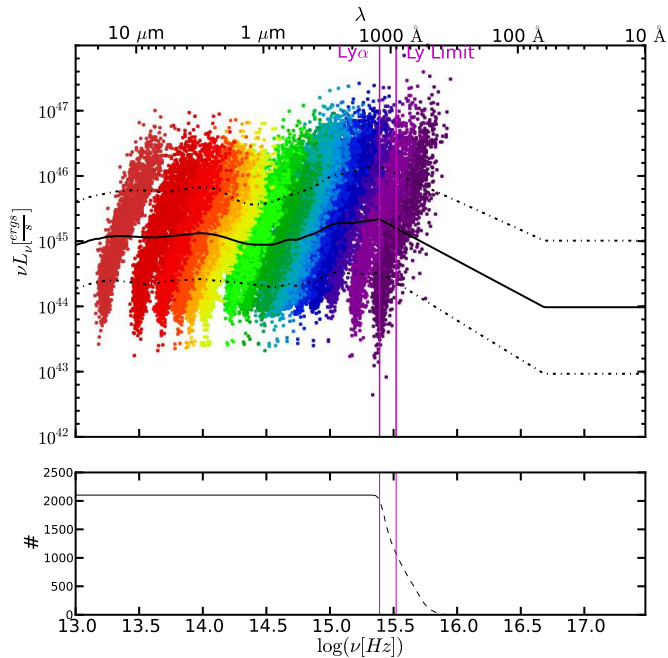}
 \caption{Top: mean ``gap filling'' SED (solid line), based on $\sim$2100 quasars with full broadband wavelength coverage.
 To avoid a bias towards high luminosity at the blue end where the number of quasars drops off, the mean is truncated at 1216\,\AA, and connected with the X-ray (see text).
 Bottom: the number of SEDs averaged at each frequency.  The dotted line indicates the data that is not used due to truncation (see text).
 The two vertical magenta lines show the positions of the Ly$\alpha$ line at 1216\,\AA\ and the Lyman limit at 912\,\AA.}
 \label{mean_full}
\end{figure}

To find the corrected mean SED, we first construct a ``gap filling'' SED by looking at the mean SED of $\sim 2100$ quasars with full wavelength coverage; i.e., they have at least four data points in mid-IR, at least three data points in the near-IR, and full coverage with {\em GALEX} in the UV. Redshift effects cause the SED coverage in the rest frame to drop off sharply around the Lyman series.  To avoid the mean being biased towards the higher luminosity SEDs in this region we have truncated this gap filling mean SED where the total number of quasars drops (around 1216\,\AA) and use the mean UV luminosity, $\langle L_{2500{\rm \AA}} \rangle$, to find the X-ray luminosity using Equation (\ref{just_eq}). Figure \ref{mean_full} shows the resulting mean SED that we will use for gap filling individual quasar SEDs that lack full coverage in all of the bandpasses considered herein.

To find the mean SED for the {\em entire} sample, we apply the corrections described in Section~\ref{corrections}, but only gap repairing filters with wavelengths $\lambda>912$\AA, so that way we avoid gap filling in a region that is not well sampled with our gap-filling SED. We then truncate the mean at 912\,\AA\ and use the mean UV luminosity, $\langle L_{2500{\rm \AA}} \rangle$, to find the X-ray luminosity using Equation~(\ref{just_eq}).
We do this instead of using actual X-ray data since there are too few X-ray detections and \citet{Steffen06} have already done the careful work of extracting the relationship between the UV and the X-ray parts of the SED.

\begin{figure}
 \centering
 \includegraphics[width=3.5in]{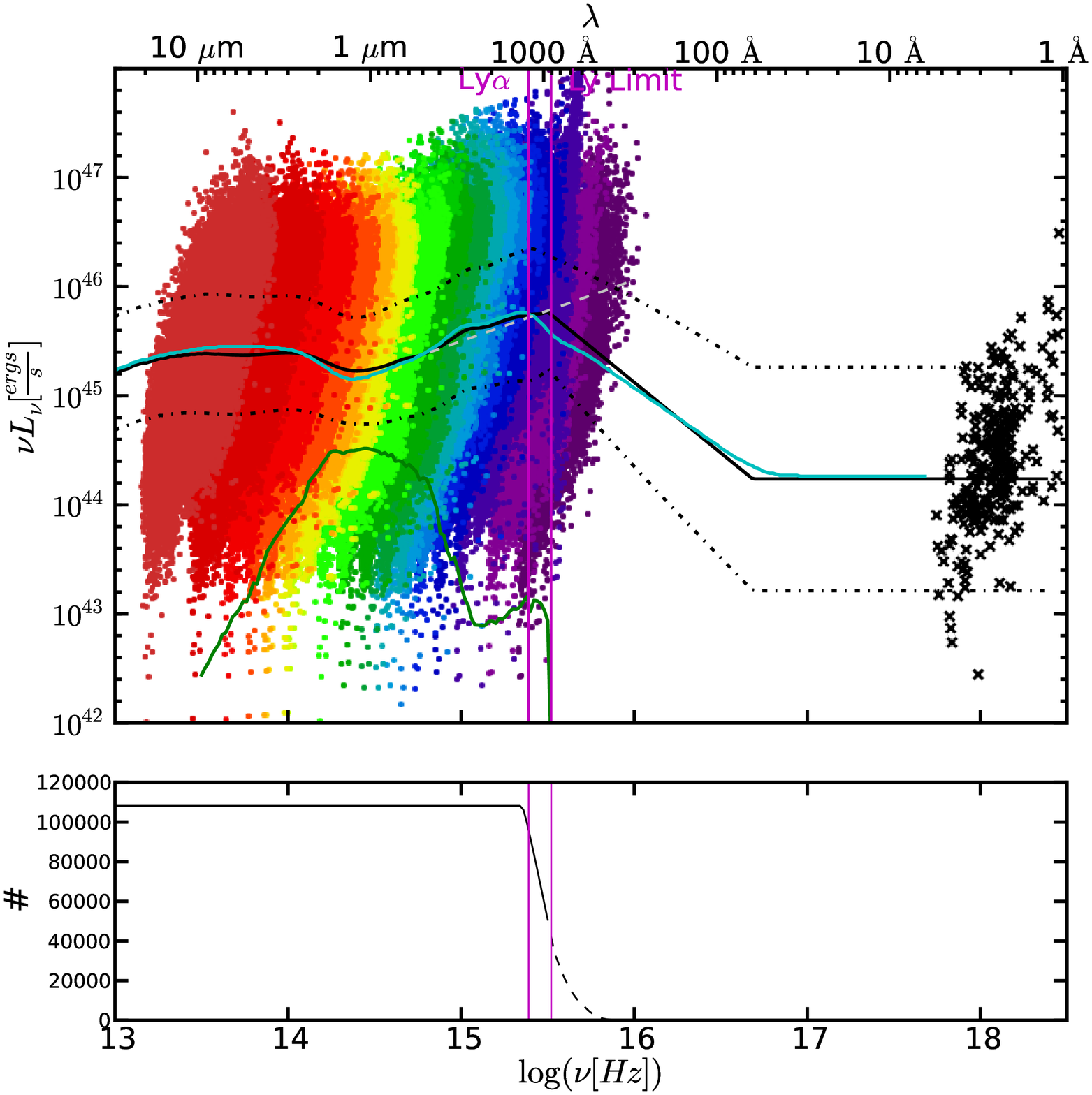}
 \caption{Top: mean corrected SED (black) and data points for 108,184 quasars with gap filling, edge filling, host galaxy removal, and spectral corrections (see text).  The mean SED from \citet[][cyan]{Richards06}, an $\alpha_{\nu}=-0.44$ power law \citep[][dashed grey]{Vanden01}, a typical elliptical host galaxy (green), and ChaMP X-ray data \citep[][black crosses]{Green09} are included for comparison. The mean has been truncated at the Lyman limit to avoid being biased to high-luminosities when the number of quasars drops off (see text).
 See legend of Figure~\ref{mean_no_cor} for color coding of scatter points.
 Bottom: the number of SEDs averaged at each frequency. The dotted line indicates where the mean SED is truncated and connected with the X-ray. The two vertical magenta lines show the positions of the Ly$\alpha$ line at 1216\,\AA\ and the Lyman limit at 912\,\AA.  
}
 \label{mean_gap_cor}
\end{figure}

Both radio-loud and radio-quiet quasars are included.  The mean SED of 108,184 quasars is given in tabular form in Table \ref{mean_sed_table}.  Figure~\ref{mean_gap_cor} shows the resulting mean with the \citet{Richards06} mean, the \citet{Vanden01} power law, a typical host galaxy, and the ChaMP X-ray data included for comparison.  The difference between our mean SED and that of  \citet{Richards06}  in the 912--1216\AA\ region is due to our more careful corrections of systematic effects in this region as described in Section~\ref{corrections}.  The abrupt change in the slope of the SED at $\sim 1100\,{\rm \AA}$ is not expected to be real, but rather represents our lack of knowledge of this region of the SED. We are largely limited to simply connecting the two better known regions of the NUV and soft X-ray with a power law. We will further discuss the range of possible FUV continua in Section \ref{EUV}.

\begin{deluxetable*}{lcccccccccccccccccc}
\tabletypesize{\scriptsize}

\tablecaption{\label{mean_sed_table} Mean Quasar SEDs }
\tablehead{ \colhead{}  &\colhead{}  &\colhead{}  &\multicolumn{2}{c}{} &\multicolumn{2}{c}{} &\multicolumn{2}{c}{} &\multicolumn{2}{c}{\ion{C}{4} Line} &\multicolumn{2}{c}{\ion{C}{4} Line} &\multicolumn{2}{c}{} &\multicolumn{2}{c}{} &\multicolumn{2}{c}{} \\
\colhead{$\log{(\nu)}$}  &\colhead{All}  &\colhead{$\sigma_{\textup{All}}$}  &\multicolumn{2}{c}{Low Lum.} &\multicolumn{2}{c}{Mid Lum.} &\multicolumn{2}{c}{High Lum.} &\multicolumn{2}{c}{Zone 1} &\multicolumn{2}{c}{Zone 2} &\multicolumn{2}{c}{UV Bump} &\multicolumn{2}{c}{Scott UV} &\multicolumn{2}{c}{Casebeer UV}  }
\startdata
13.0&  45.201&  0.52&  44.844&  0.24&  45.272&  0.2&  45.632&  0.25&  45.434&  0.31&  45.831&  0.33&  45.213&  0.55&  45.248&  0.55&  45.248&  0.55\\
13.02&  45.221&  0.52&  44.864&  0.24&  45.292&  0.2&  45.652&  0.25&  45.454&  0.31&  45.851&  0.33&  45.233&  0.55&  45.268&  0.55&  45.268&  0.55\\
13.04&  45.231&  0.52&  44.874&  0.24&  45.302&  0.2&  45.662&  0.25&  45.464&  0.31&  45.861&  0.33&  45.243&  0.55&  45.278&  0.55&  45.278&  0.55\\
13.06&  45.241&  0.52&  44.884&  0.24&  45.312&  0.2&  45.672&  0.25&  45.474&  0.31&  45.871&  0.33&  45.253&  0.55&  45.288&  0.55&  45.288&  0.55\\
13.08&  45.251&  0.52&  44.894&  0.24&  45.322&  0.2&  45.682&  0.25&  45.484&  0.31&  45.881&  0.33&  45.263&  0.55&  45.298&  0.55&  45.298&  0.55\\
13.1&  45.271&  0.52&  44.914&  0.24&  45.342&  0.2&  45.702&  0.25&  45.504&  0.31&  45.901&  0.33&  45.283&  0.55&  45.318&  0.55&  45.318&  0.55\\
13.12&  45.281&  0.52&  44.924&  0.24&  45.352&  0.2&  45.712&  0.25&  45.514&  0.31&  45.911&  0.33&  45.293&  0.55&  45.328&  0.55&  45.328&  0.55\\
13.14&  45.291&  0.52&  44.934&  0.24&  45.362&  0.2&  45.722&  0.25&  45.524&  0.31&  45.921&  0.33&  45.303&  0.55&  45.338&  0.55&  45.338&  0.55\\
13.16&  45.301&  0.52&  44.944&  0.24&  45.372&  0.2&  45.732&  0.25&  45.534&  0.31&  45.931&  0.33&  45.313&  0.55&  45.348&  0.55&  45.348&  0.55\\
13.18&  45.301&  0.52&  44.944&  0.24&  45.372&  0.2&  45.732&  0.25&  45.534&  0.31&  45.931&  0.33&  45.313&  0.55&  45.348&  0.55&  45.348&  0.55
\enddata

\tablecomments{All of the SEDs are taken to have $\alpha_x=-1$ above 0.2 keV. Units are log(erg s$^{-1}$). 
This table is published in its entirety in the electronic edition of the online journal. A portion is shown here for guidance regarding its form and content.}
\end{deluxetable*}

\subsection{Sub-sampled Mean SEDs}

While the overall mean quasar SED is a useful tool, examining how the SED changes as a function of various quasar parameters may shed light on the physical processes of the central engine.  For example, \citet{Richards06} found that bolometric corrections (see Section~\ref{bol}) differed by as much as a factor of two in the extremes of quasar types, but with only 259 objects \citet{Richards06} did not have the data necessary to comment on what physics is behind the range of bolometric corrections.
\citet{Marconi04} provide a luminosity-dependent bolometric correction, but this is largely dependent on the  \luvaox\ relationship and is built into the \citet{Richards06} bolometric corrections.  What we seek (with the aid of a substantially larger sample) is a more physical understanding of the SED differences and the resulting changes in bolometric correction.  

In an attempt to better understand the physics that lead to differences in SEDs (and bolometric corrections), we consider two parameters herein, specifically the UV luminosity and \civ\ emission line properties.  While it is also important to consider SEDs as a function of mass and accretion rate, those quantities are not directly measurable; we will leave that analysis to future work (C. M. Krawczyk et al. 2013 in preparation).
Luminosity-dependent SEDs (Section~\ref{Luminosity SED}) are of interest because of the known \luvaox\ relationship and strong dependence of accretion disk wind physics on the UV to X-ray flux ratio \citep[e.g.][]{Proga00}.  Examining the mean SED as a function of \civ\ emission line properties (Section~\ref{CIV SED}) is interesting because that line may be an indicator of the true SED as it serves as a diagnostic of which two components dominate the broad emission line region \citep[BELR;][]{Richards11,Wang11}.
In fact, UV emission lines like \civ, with ionization potentials in the EUV part of the spectrum, may even be an indicator of the unseen EUV part of the SED (see Section~\ref{EUV}).

\subsubsection{Luminosity-dependent Mean} \label{Luminosity SED}

\begin{figure}
 \centering
 \includegraphics[width=3.5in]{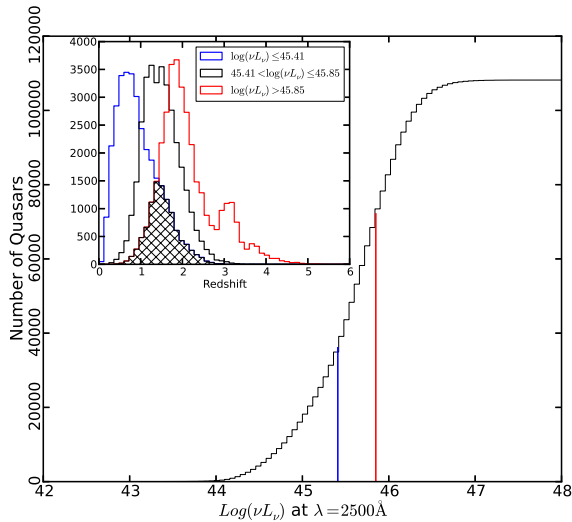}
 \caption{Cumulative histogram of $\log{(\nu L_{\nu})}\mid_{\lambda=2500 {\rm \AA}}$ for all non-reddened quasars. Vertical lines show luminosity cuts that result in three equally populated bins. The inset shows the redshift distribution for these three bins, with the overlap shaded in.}
 \label{cum_hist}
\end{figure}

The well-known nonlinear relationship between the UV and X-ray luminosities means that the SED (and thus the bolometric correction) {\em must} be a function of luminosity.  As a first test, we have split the sample into three equally populated luminosity bins, each containing 36,061
quasars.  Figure \ref{cum_hist} shows the cumulative histogram of $\log{(\nu L_{\nu})}\mid_{\lambda=2500{\rm \AA}}$ for all 
quasars, with the vertical lines showing the luminosity cuts.
In order to separate the changes that potentially arise from evolution from those that are luminosity-dependent, we marginalize over redshift by selecting sub-samples from each bin that each have the same redshift distribution.
The inset of Figure~\ref{cum_hist} shows the redshift distribution for each of the bins; the shaded region shows the redshift distribution of the sub-samples.  This process leaves $\sim$10,000 quasars in each luminosity bin as compared to the 259 {\em total} objects in \citet{Richards06} across all luminosity bins.  

\begin{figure*}
 \centering
 
\begin{tabular}{cc}
\includegraphics[width=3.2in]{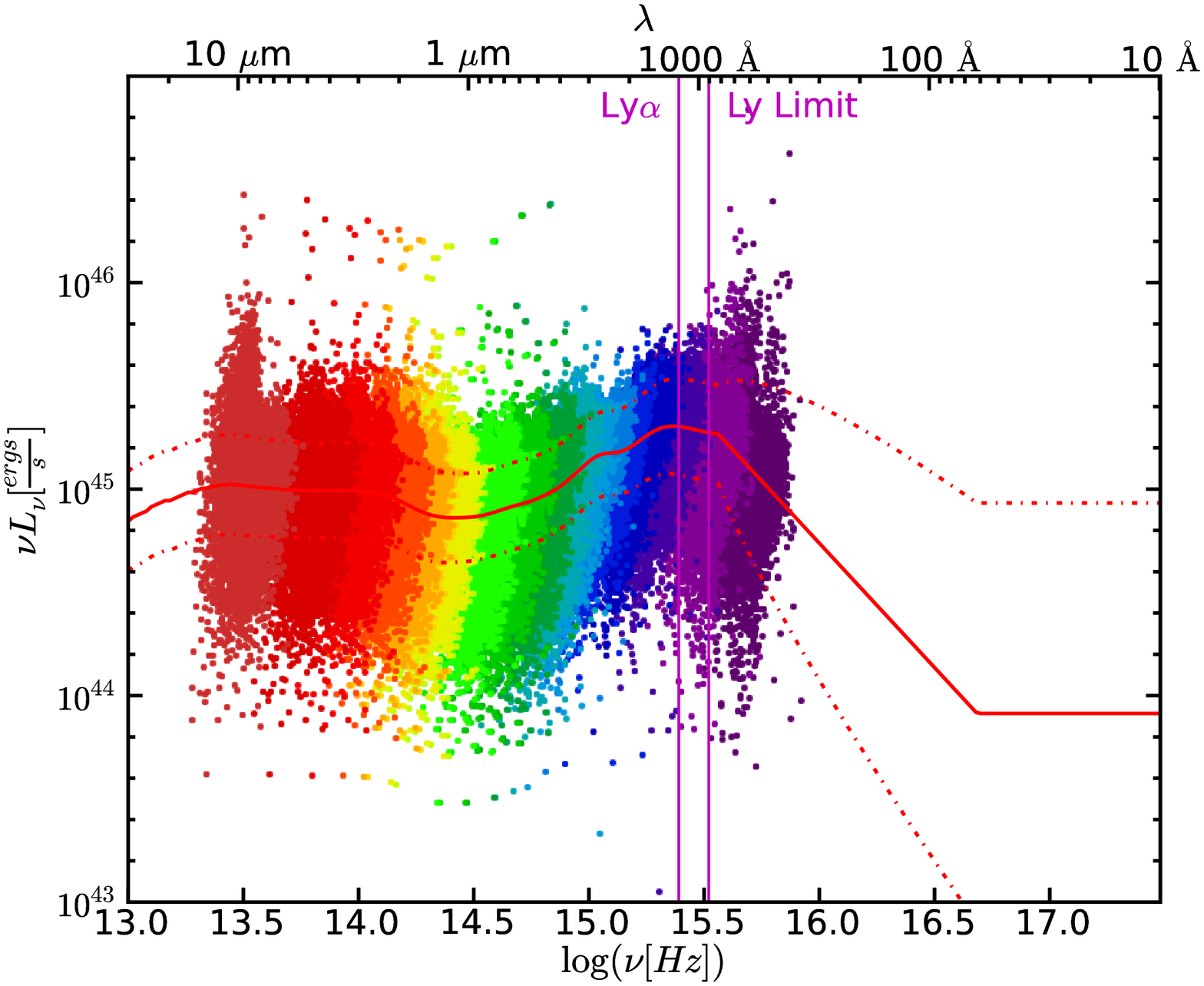} & \includegraphics[width=3.2in]{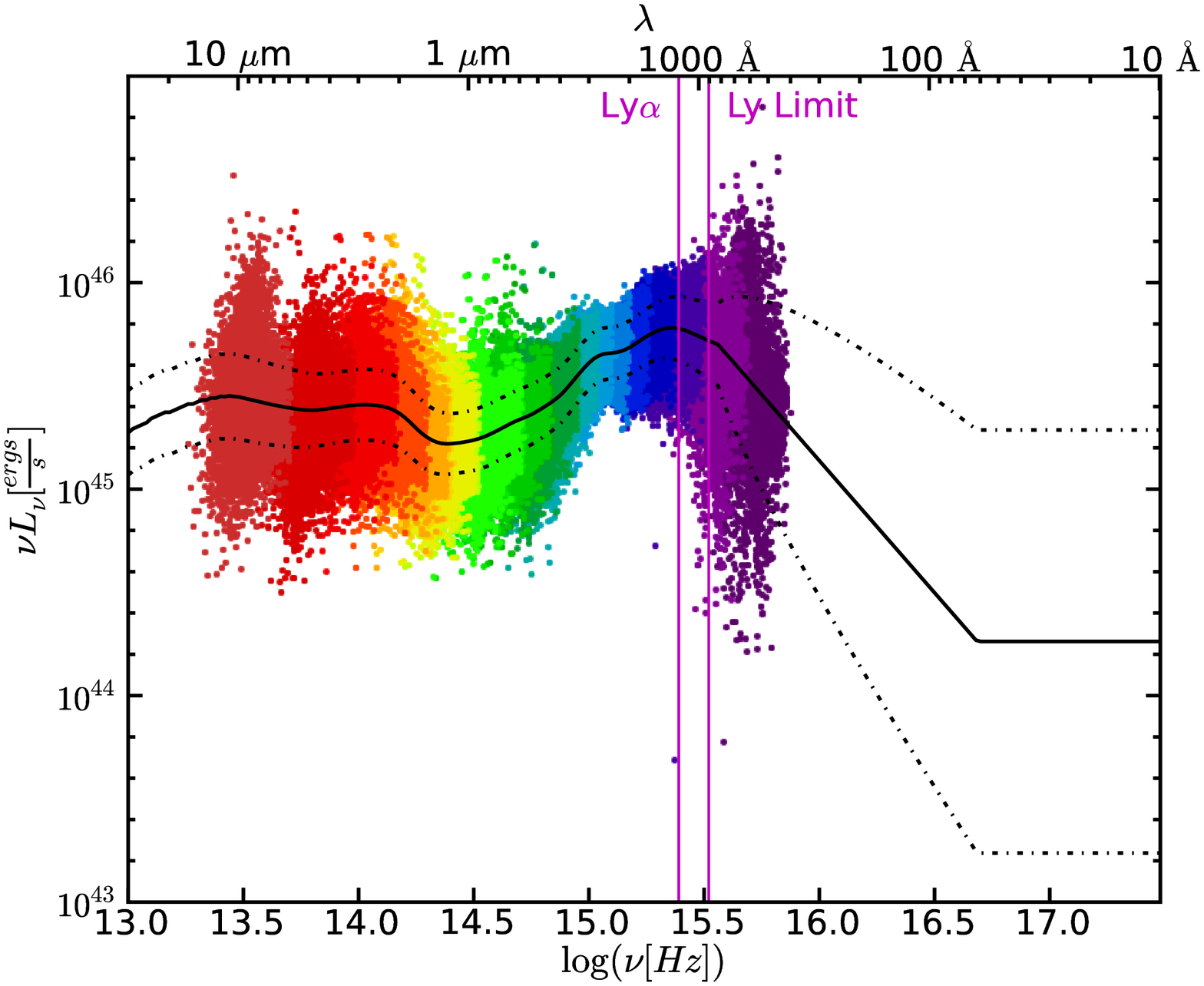} \\
 (a) Low luminosity SEDs & (b) Mid luminosity SEDs \\
\includegraphics[width=3.2in]{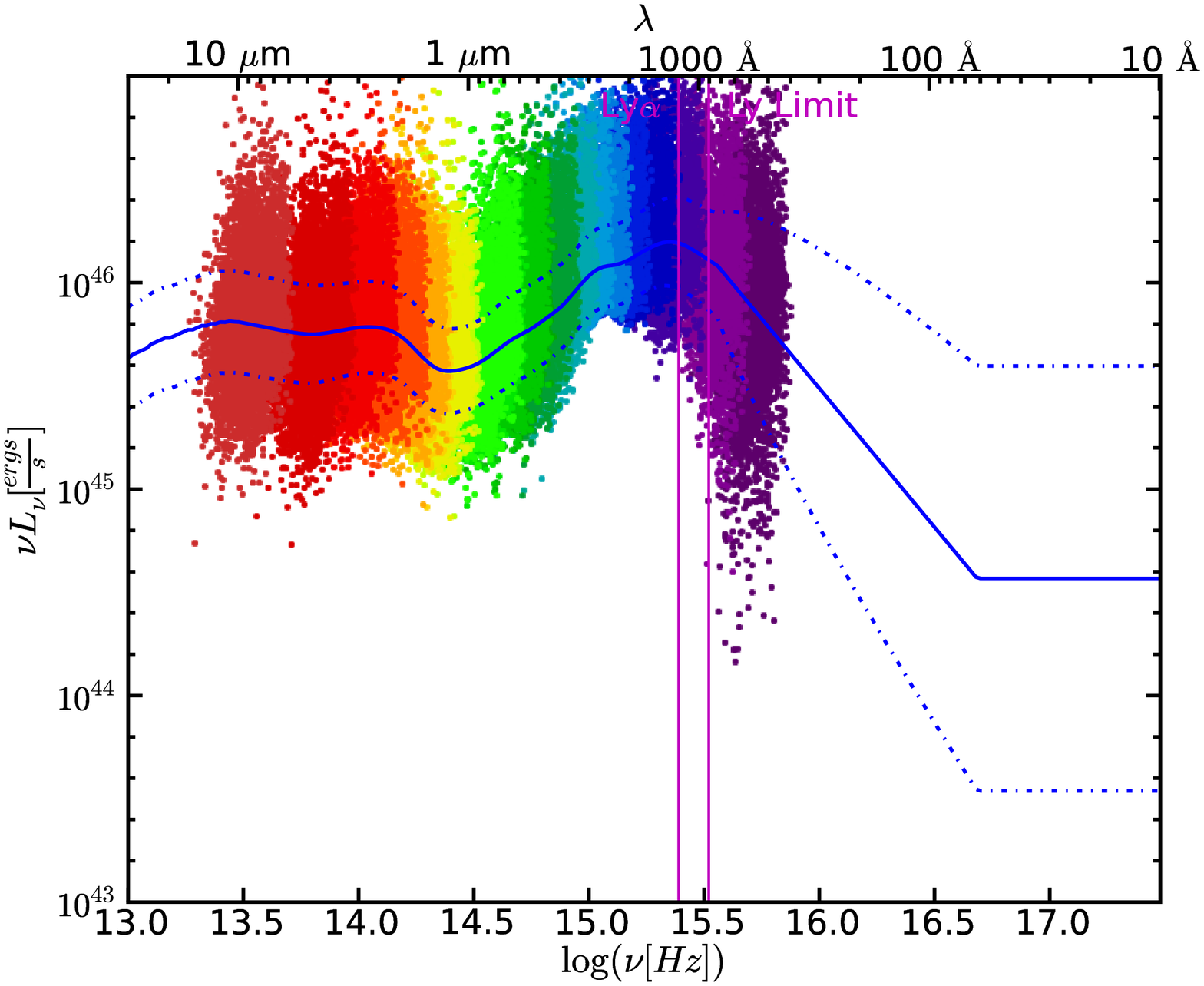} & \includegraphics[width=3.2in]{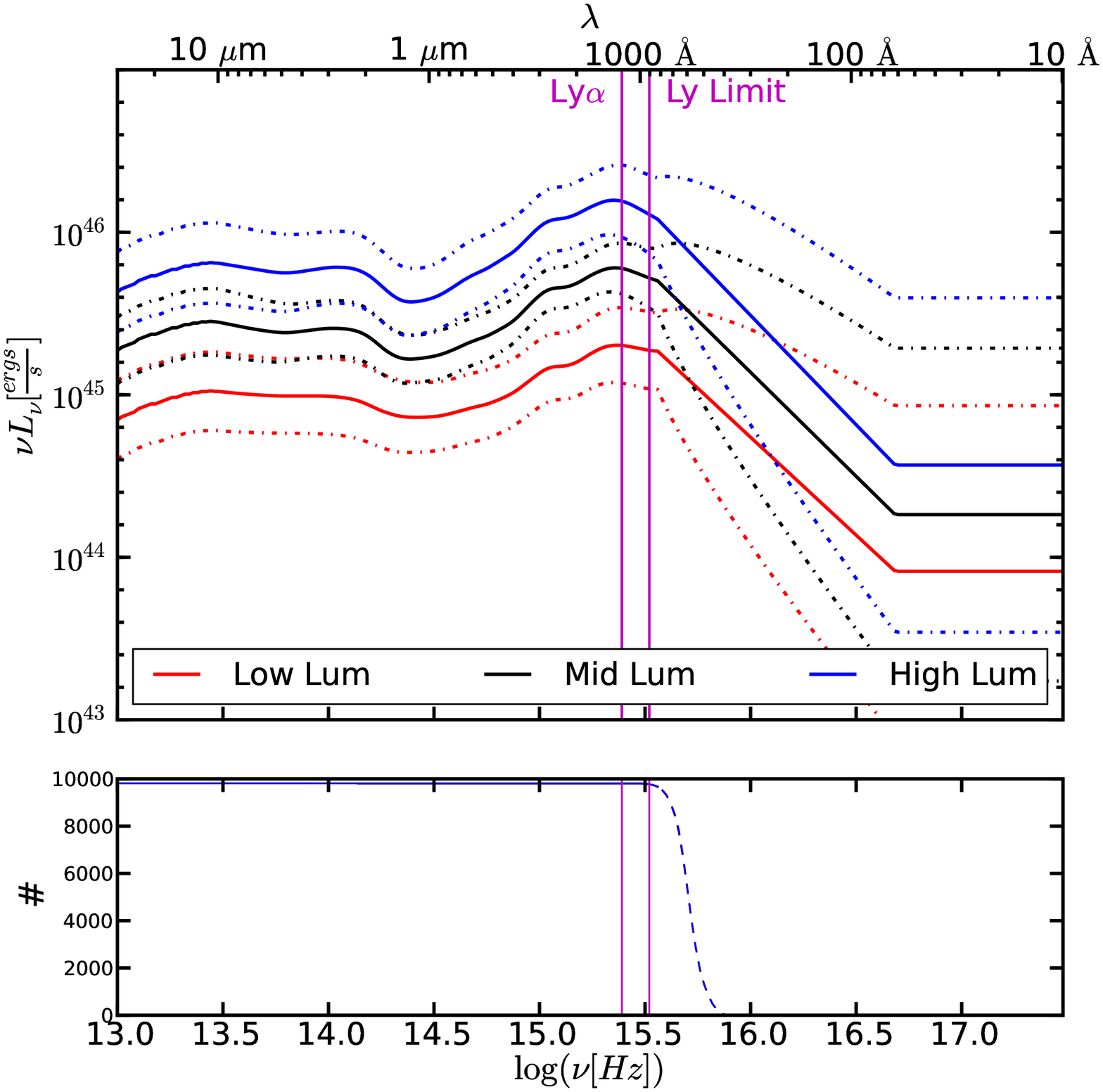} \\
 (c) High luminosity SEDs & (d) Luminosity dependent SEDs
 \end{tabular}

 \caption{Mean SEDs for low ($\log{(\nu L_{\nu})}\mid_{\lambda=2500 {\rm \AA}} \leq 45.41$) luminosity quasars (a), mid ($45.41< \log{(\nu L_{\nu})}\mid_{\lambda=2500 {\rm \AA}} \leq 45.85$) (b), and high ($45.85 < \log{(\nu L_{\nu})}\mid_{\lambda=2500 {\rm \AA}}$) (c). Panel (d) shows the three curves overplotted.
 See Figure~\ref{mean_no_cor} for an explanation of the scatter points in panels (a)-(c) and the bottom histogram of panel (d). 
 The high luminosity SED has more hot dust emission (at 2--4$\mu$m), a smaller Balmer continuum ($\lambda<3000{\rm \AA}$), a harder (bluer) optical spectrum ($\lambda<3000{\rm \AA}$), and a softer (redder) UV spectrum ($\lambda>3000{\rm \AA}$); see Figure~\ref{norm_seds}(a) and (b).}
 \label{lum_bin}
\end{figure*}

We calculated the luminosity-dependent mean SEDs in an iterative way.  As a first step we follow the same steps as the overall mean SED: we used our gap-filling SED to interpolate/extrapolate over gaps in the photometry with $\lambda>912$\,\AA\ and truncated the mean SEDs at 912\AA.  These mean SEDs were then connected to the X-ray points determined using the mean UV luminosity for each sample.  The resulting mean SEDs were then used as the new gap-filling SEDs for each luminosity bin.  For all remaining iterations all missing photometry was gap filled, and the resulting mean SEDs were truncated at 800\,\AA\ (where the sampled number starts to fall) before connection to the X-ray. Figure~\ref{lum_bin} shows the mean SEDs and data for each of the luminosity bins after 10 iterations.  
The iteration process described above ensures that our mean SED follows the quasars with data in each luminosity bin instead of the initial gap filling model.  The redshift marginalization process means that the mean redshift of all three luminosity samples is $z \sim 1.5$.  The mean SEDs are given in tabular form in Table \ref{mean_sed_table}.

\begin{figure*}
 \centering
 \begin{tabular}{cc}
\includegraphics[width=3.2in]{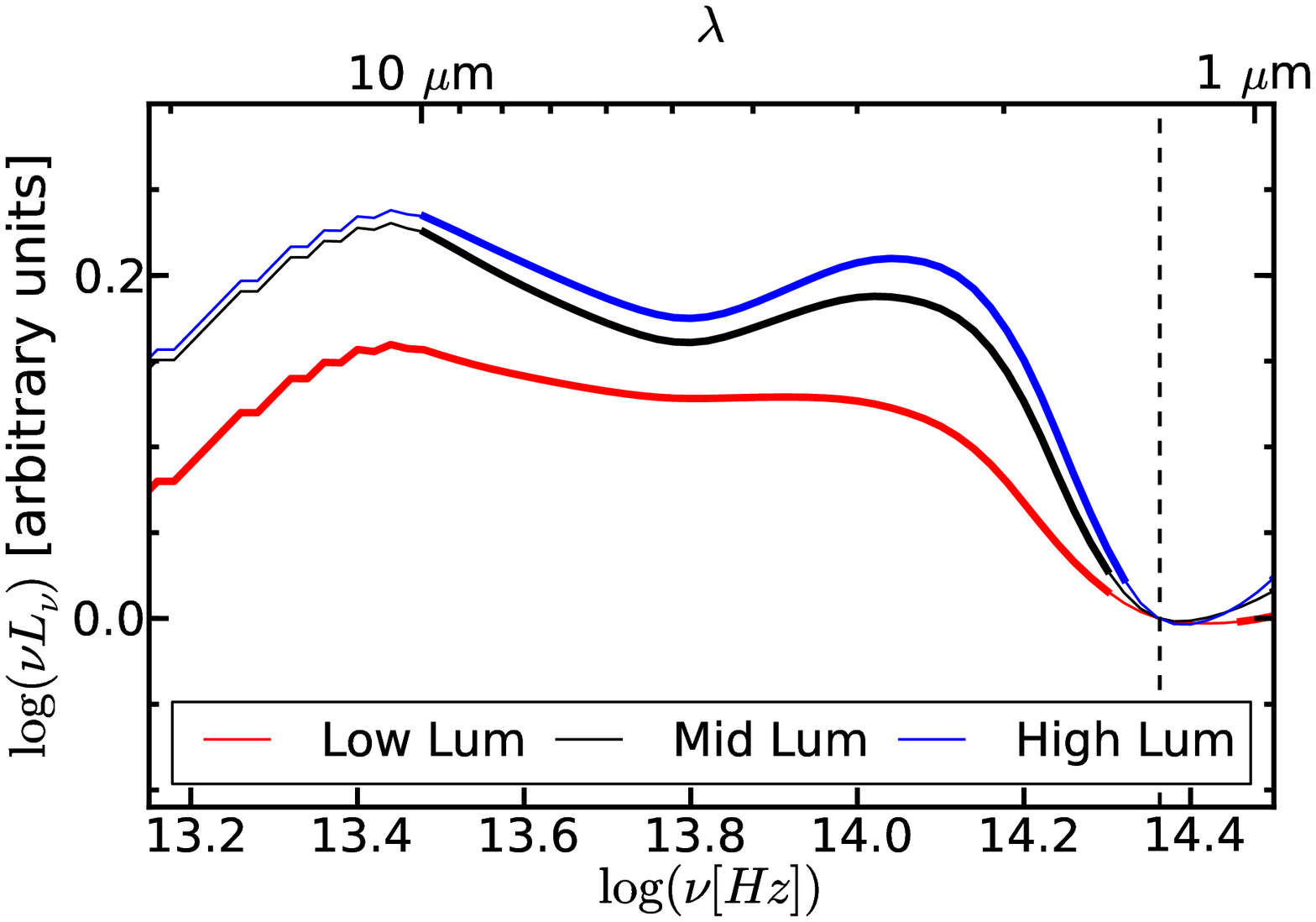} & \includegraphics[width=3.2in]{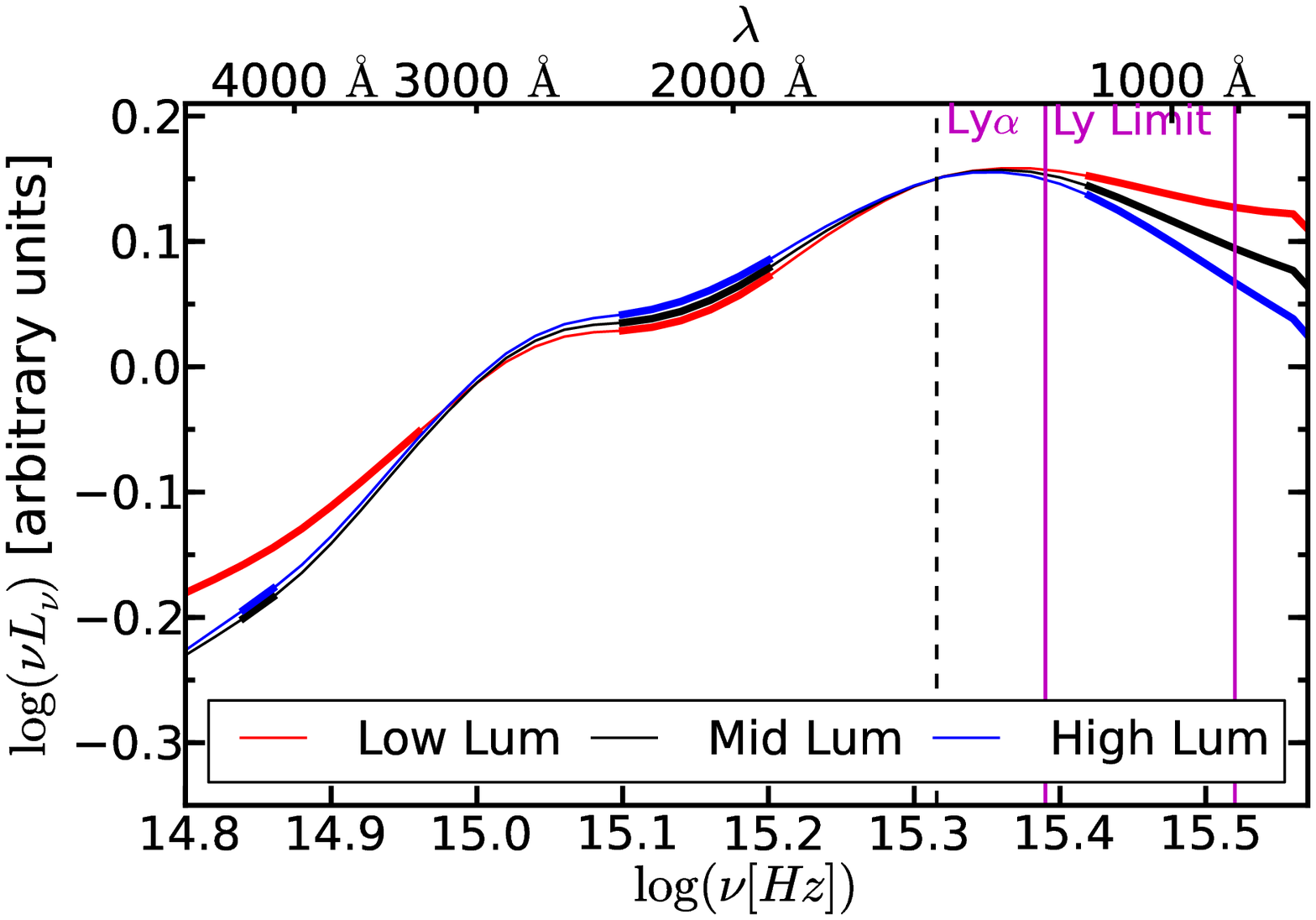} \\
(a) Luminosity SEDs normalized at 1.3\,$\mu$m & (b) Luminosity SEDs normalized at 1450\,\AA \\
\includegraphics[width=3.2in]{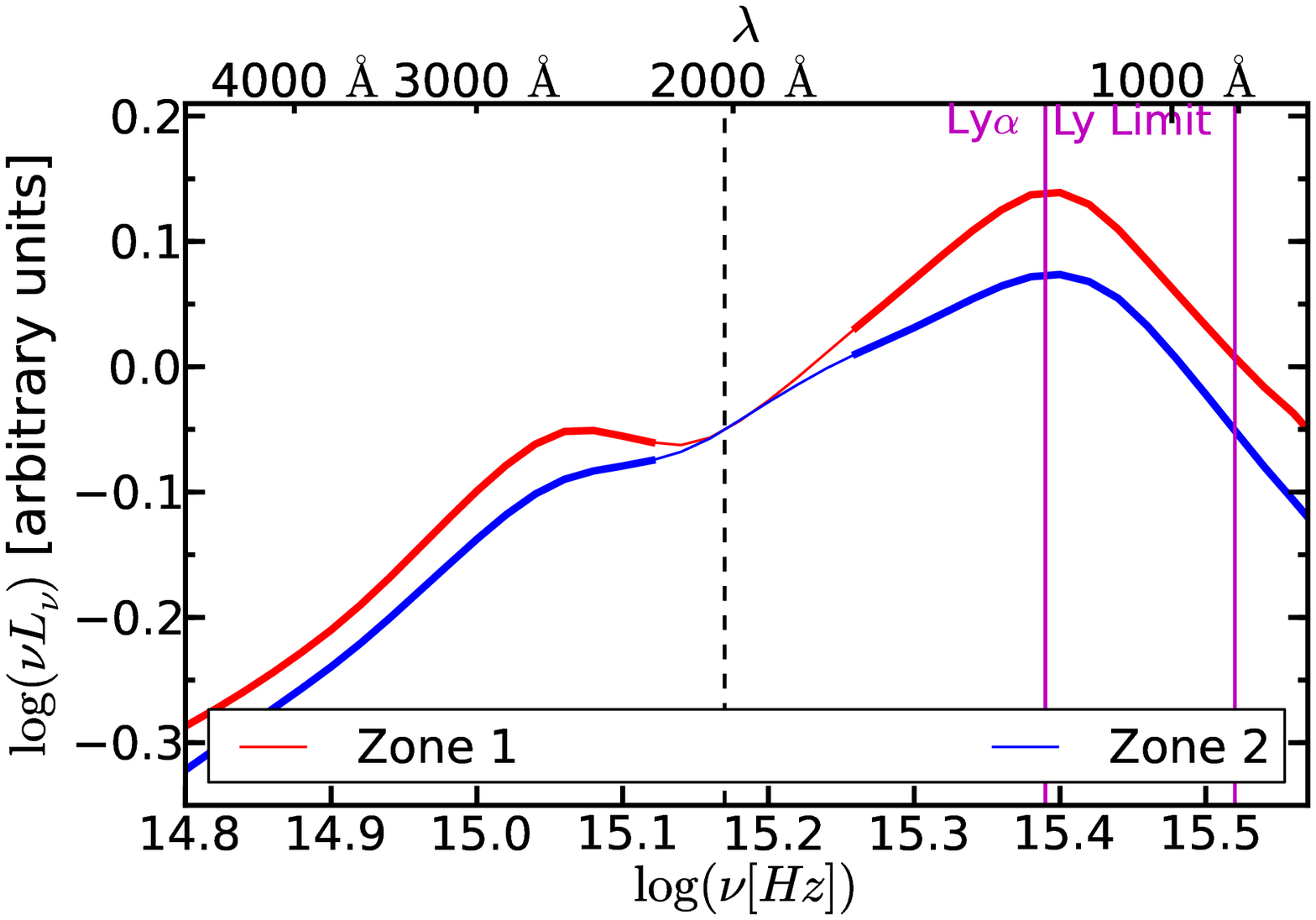} & \includegraphics[width=3.2in]{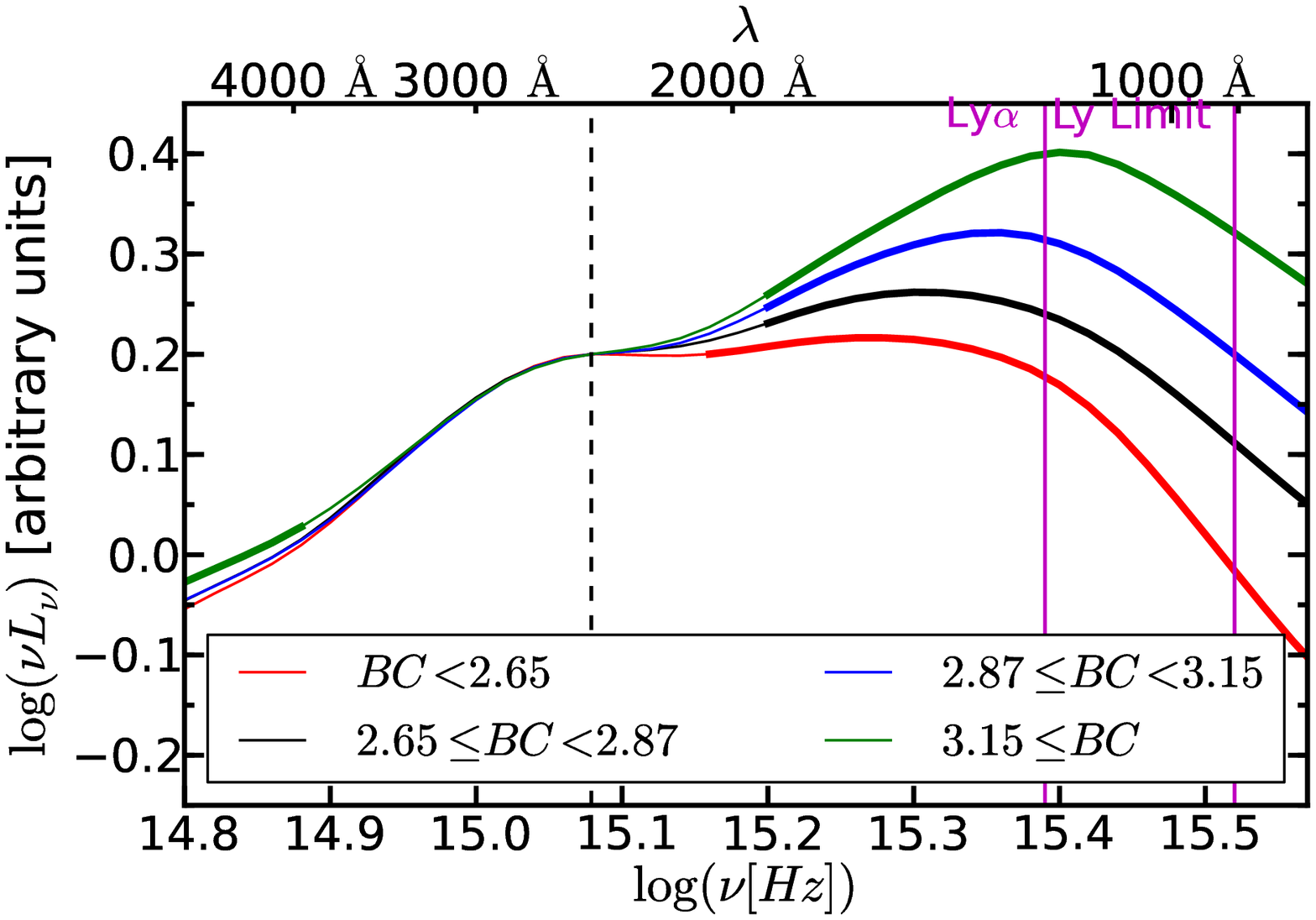} \\
(c) \civ\ SEDs normalized at 2027\,\AA & (d) BC SEDs normalized at 2500\,\AA
\end{tabular}

 \caption{Normalized SEDs with the $y$-axis showing relative luminosity. The thick lines indicate where the probability of the mean SEDs being the same is less than 1\% and the vertical dashed lines indicate the normalization frequency. (a) Luminosity-dependent SEDs normalized at 1.3\,$\mu$m; high-luminosity quasars appear to have more hot dust emission.  (b) Luminosity-dependent SEDs normalized at 1450\,\AA; high-luminosity quasars have softer EUV continua. (c) \civ-dependent SEDs normalized at 2027\,\AA. Zone 1 (disk-dominated) quasars have more Balmer continuum and stronger Ly$\alpha$ emission then Zone 2 (wind-dominated) quasars. See Figure~\ref{civ_bin}. (d) Bolometric correction-dependent SEDs normalized at 2500\,\AA, showing the effect that most of the range in BC comes from the range of UV and EUV continua (and dust). 
}
 \label{norm_seds}
\end{figure*}

Figure~\ref{norm_seds}(a) shows the mid-IR region of the three SEDs when they are normalized at 1.3$\mu$m and Figure~\ref{norm_seds}(b) shows the optical ($\lambda<3000{\rm \AA}$) and UV region ($\lambda>3000{\rm \AA}$) when normalized at 1450\AA. The thick lines indicate where a Welch's $t$ test shows the mean SEDs have less than a 1\% chance of being the same ($p<$0.01).  The high-luminosity SED has more hot dust emission (at 2--4$\mu$m), a small Balmer continuum ($\lambda<3000{\rm \AA}$), a harder (bluer) optical spectrum, and a softer (redder) UV spectrum.
The low-luminosity quasars would have to have host galaxies $\sim$8 times more luminous than assumed in order for the optical slopes at $\lambda<3000{\rm \AA}$ to agree between the low-luminosity and high-luminosity SEDs (thus the redder continuum is likely to be intrinsic and not due to host-galaxy contamination).  The UV region is sensitive to the Lyman series corrections (Section~\ref{sec:emission}), but these corrections are not luminosity dependent.  The Welch's $t$ test shows the prominent \tenmum\ silicate bumps in the high-luminosity objects are statistically significant.
The IR bumps are discussed in more detail in \citet{Gallagher07b}. 

One of our goals in considering the luminosity-dependent mean is to investigate the shape of the SEDs over $500<\lambda_{{\rm rest}}\,({\rm \AA}) <1200$.  
Given the corrections necessary in this region (Section~\ref{sec:emission}), the shape here is necessarily uncertain; however, the data supports a harder slope with lower luminosity
(see Figure~\ref{norm_seds}(b)).  This is consistent with the results of \citet{Scott04} and suggests that the X-ray flux may not be the only part of the SED that has a nonlinear relationship to \ltwofive.  This behavior is also seen in the spectral Principal Component Analysis (PCA) of \citet{Yip04} who found that their third eigenvector shows an anti-correlation of the continua on either side of Ly$\alpha$.  In short, quasars that are bluer longward of Ly$\alpha$ have softer (redder) continua in the Ly$\alpha$ forest region.   This can have important consequences for bolometric corrections; see Section~\ref{bol}.

\subsubsection{\civ-dependent Mean} \label{CIV SED}

While the Baldwin Effect \citep{Baldwin77} reveals that there is a relationship between the strength of UV emission lines and the continuum luminosity, \citet{Richards11} have argued that the \civ\ (and other UV emission lines) properties are better diagnostics of the {\em shape} of the SED than its absolute scaling.  If that is the case, SEDs made as a function of emission line properties such as \civ\ blueshift and EW \citep{Richards11} or PCA \citep[e.g., ``Eigenvector 1'';][]{Boroson92,Brotherton99} may reveal interesting differences.  Any such differences would have important implications for bolometric corrections of individual objects.

\begin{figure*}
 \centering
  \begin{tabular}{cc}
\includegraphics[width=3.2in]{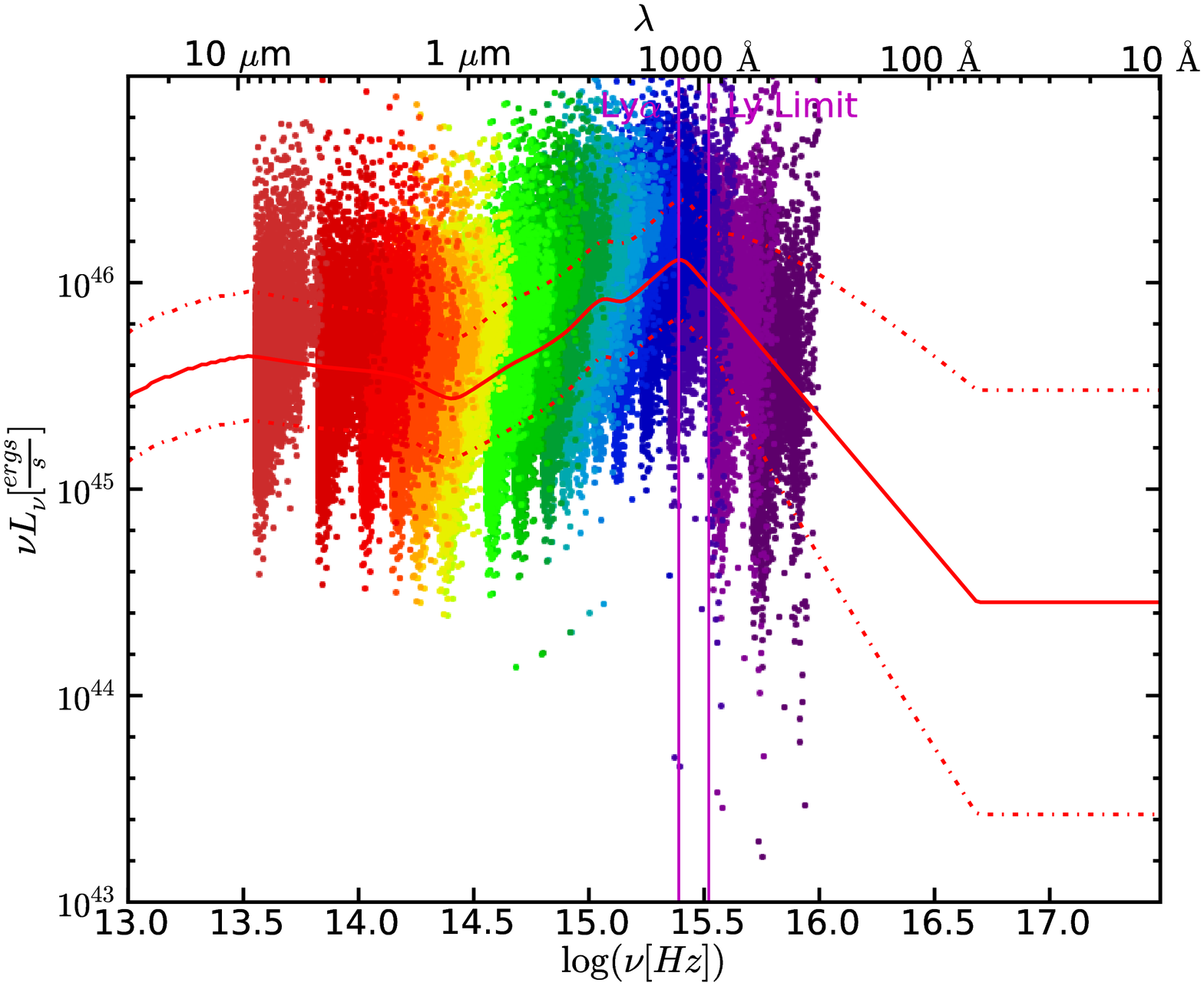} & \includegraphics[width=3.2in]{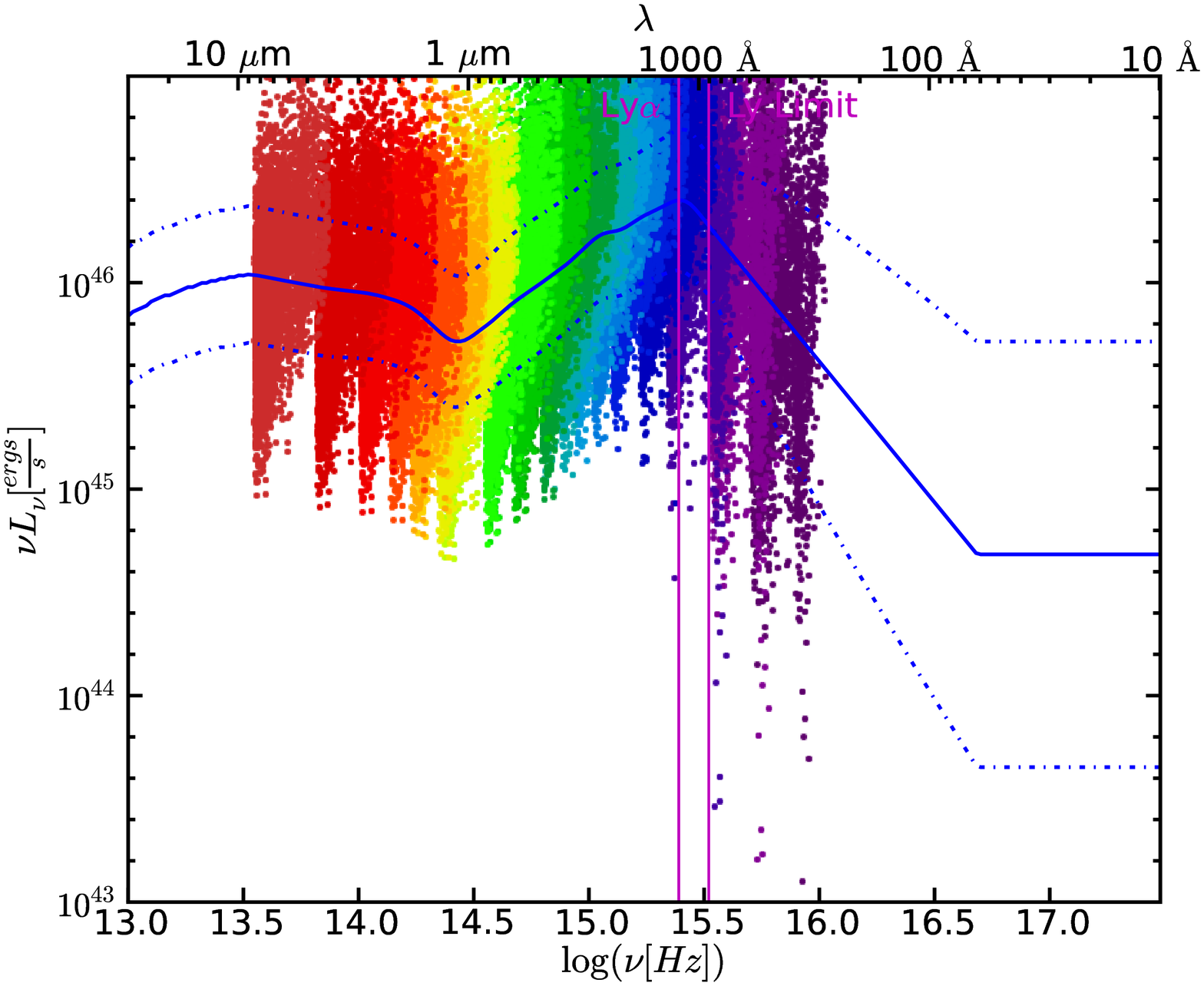} \\
 \\
 (a) Zone 1 SEDs & (b) Zone 2 SEDs \\
 \end{tabular}
 \begin{tabular}{c} 
\includegraphics[width=3.2in]{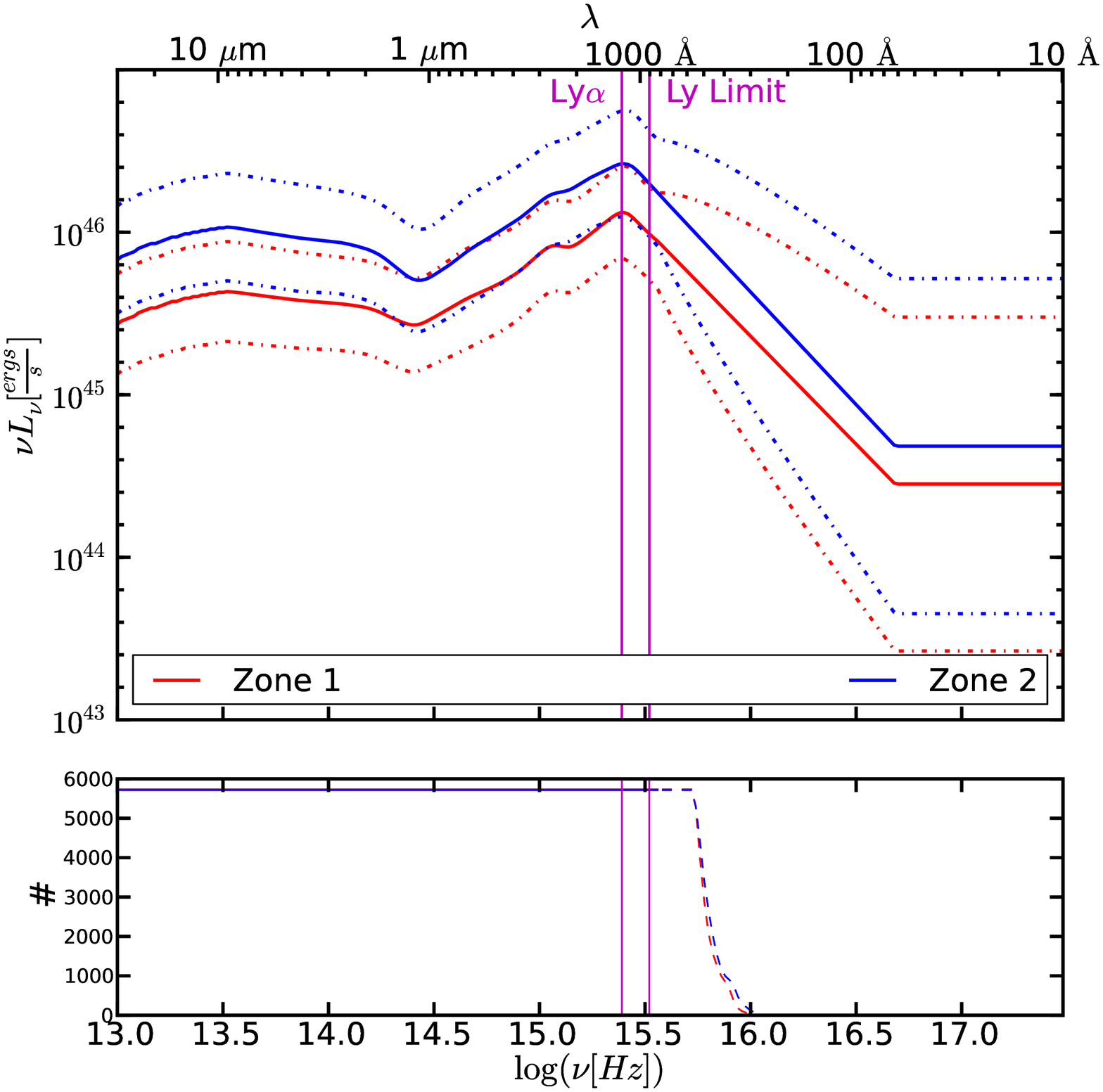} \\
 (c) \ion{C}{4} dependent SEDs
 \end{tabular}
 
 \caption{Mean SEDs for quasars in two zones of the \ion{C}{4} blueshift vs. \ion{C}{4} EW plane.  Panel (a) shows Zone 1: BS $< 600$ km s$^{-1}$ and EW $> 50 \mbox{ \AA}$ \citep[disk-dominated objects in ][]{Richards11}. Panel (b) shows Zone 2: BS $> 1200$ km s$^{-1}$and EW $< 32 \mbox{ \AA}$ \citep[wind-dominated objects in ][]{Richards11}. Panel (c) shows the two curves overplotted.
 See Figure~\ref{mean_no_cor} for explanation of scatter points in panels (a and b) and the bottom histogram of panel (c). 
 The zone 1 mean SED shows a larger Balmer continuum ($\lambda<3000{\rm \AA}$) and stronger Ly$\alpha$ emission, consistent with eigenvector 1 analyses.
 Figure~\ref{norm_seds} shows an expanded version of panel (c).}
 \label{civ_bin}
\end{figure*}

Thus, in addition to luminosity sub-samples, we have also divided the data as a function of \civ\ blueshift and rest-frame EW \citep{Richards11}.  To see if the SED shape depends on these properties, we have taken two zones that are representative of extrema in disk-wind structures according to \citet[][see also \citet{Wang11}]{Richards11}
:  Zone 1, blueshifts $<$ 600 km s$^{-1}$ and EW $>$ 50 \AA\ (i.e., with ``disk-dominated'' BELRs) and Zone 2, blueshifts $>$ 1200 km s$^{-1}$ and EW $<$ 32 \AA\ (i.e., with ``wind-dominated'' BELRs). 
These cuts are chosen to have roughly the same number of objects in each sample.
In Zone 1 there are 5736 quasars with a mean $\log{(\nu L_{\nu})}\mid_{\lambda=2500{\rm \AA}}=45.9$, and Zone 2 contains 5713 quasars with a mean $\log{(\nu L_{\nu})}\mid_{\lambda=2500{\rm \AA}}=46.2$.
Figure \ref{civ_bin} shows the data and mean SEDs for the quasars in each zone.  For these mean SEDs, we used the mean mid-luminosity SED for the initial gap repair to {\em all} the filters and truncated the mean at 800\AA\ before connecting it with the X-ray.  We then took the resulting mean SED and used that to gap repair in the same way; this was repeated 10 times.  We note that there is a difference in mean luminosity between Zones 1 and 2; however, we have not gap-filled with the luminosity-dependent SEDs in order to isolate any residual differences. The mean SEDs are given in tabular form in Table \ref{mean_sed_table}.

Our primary interest in these SEDs lies in the UV part of the spectrum; however, it seems that there is relatively little that can be learned about that region from this sample.  Seeing \civ\ requires that the sample is restricted to $z\gtrsim1.6$, which has a number of consequences.  In particular, it limits the sample to the highest luminosities and the redshift limit introduces a bias toward LLS \citep{Worseck11}, which may be the cause of the steep fall-off beyond Ly$\alpha$.  
However we are able to see differences in the optical/UV continuum consistent with eigenvector 1 analyses \citep{Brotherton99}.  
In particular, Figure~\ref{norm_seds}(c) shows that the Zone 1 mean SED has 
a larger Balmer continuum ($\lambda<3000{\rm \AA}$) and stronger Ly$\alpha$ emission (the difference in slopes is consistent with being due to stronger \civ\ and Ly$\alpha$ in Zone 1 mean SED relative to Zone 2 since we have only corrected for the mean emission line contribution).
We will further consider differences in the bolometric correction and what can be learned about the shape of the SED as a function of the UV emission line properties in Section~\ref{bol}, 
further investigations of the UV continuum of quasars as a function of \civ\ emission line properties are certainly warranted.

\section{Bolometric corrections} \label{bol}

One of the main goals of this study is to characterize the bolometric luminosities of quasars.  The bolometric luminosity is the integrated area under the SED curve and can be calculated as:
\begin{equation}
 L_{{\rm bol}} = \int_{0}^{\infty} L_{\nu} d\nu = \int_{-\infty}^{\infty} \ln(10) \nu L_{\nu} d\log(\nu)
\end{equation}
where Figures \ref{mean_no_cor} and \ref{mean_gap_cor} plot the quantity $\nu L_{\nu}$.
Normally, we do not have an accurate measurement of the full SED for a quasar, but we can estimate the bolometric luminosity by assuming some SED template and using a known monochromatic luminosity, $L_{\nu}$.  This ``bolometric correction'' (BC) is given as:
\begin{equation}
 {\rm BC}_{\nu} = \frac{L_{{\rm bol}}}{\nu L_{\nu}}.
\end{equation}

In the next sections we will discuss our choices for limits of integration, normalization wavelength/frequency, and SED models.

\subsection{Limits of Integration}

As was discussed by \citet{Marconi04}, the full observed SED includes
light that is not directed to the observer along its original line of
sight.  Thus, the SED determined in this manner is not the {\em
  intrinsic} one.  For example, most of the IR radiation ($\sim$\thirtymum --\onemum) is believed
to be produced in a large toroidal dusty region beyond the accretion
disk \citep[e.g.,][]{Krolik88,Elitzur06}.  The radiation from this ``dusty
torus'' arises from the reprocessing of higher energy photons emitted
by the accretion disk and re-radiated by the torus in the IR.  
Any 
optical-UV radiation that does not come directly to the observer, but is instead reprocessed by dust, is effectively double-counted when determining the true bolometric luminosity.
Because of this effect,
\citet{Marconi04} do not include the IR bump when computing their
bolometric corrections.  Their choice of integration limits of \onemum --500\,keV is well
justified, although strictly speaking the optical/UV bump must also be
corrected for dust reddening along the direct line of sight;
otherwise, the observed luminosity will be smaller than the true intrinsic luminosity.

Here we describe a similar effect in the hard X-ray band (energies larger than \twokev).  While it is thought that IR flux can include reprocessed disk emission, so too can the hard X-ray emission.  The accretion disk itself can emit thermal soft X-ray photons from the inner part of the accretion disk, but hard X-ray photons are believed to come from Compton upscattering of accretion disk photons off hot electrons in the so-called ``corona.''  Depending on the geometry of the corona, it may or may not be appropriate to include the hard X-ray part of the SED in the bolometric luminosity in the same way that \citet{Marconi04} suggest avoiding the IR emission.  If the corona is a hot spherical region surrounding the black hole \citep{Sobolewska04a} then one should exclude the highest energy photons from the tabulation of the ``intrinsic'' luminosity.  However, if the corona is more like a patchy skin to the accretion disk as in \citet{Sobolewska04b}, including the hard X-ray flux in the SED would be more appropriate.  Our solution here is an agnostic one; instead of presenting the full integrated luminosity, we will give it in four pieces: \thirtymum--\onemum, \onemum--\twokev, \twokev\--\tenkev, and \tenkev--500\,keV (see Table~\ref{BC_table}), allowing the user to determine which approach is best. 
Although, given the small amount of data and the large uncertainties in the X-ray, we do not recommend taking BCs based on monochromatic X-ray luminosities as proposed by \citet{Vasudevan07}.

\begin{deluxetable*}{lcccccccccccccc}
\tabletypesize{\footnotesize}
\tablewidth{0pt}

\tablecaption{\label{BC_table} Bolometric Luminosities and Bolometric Corrections }
\tablehead{ \colhead{SDSS ID}  &\multicolumn{2}{c}{$\log{(\nu L_{\rm 2500\,\AA})}$} &\multicolumn{2}{c}{$\log{(\nu L_{\rm 5100\,\AA})}$} &\multicolumn{2}{c}{BC$_{\rm disk}$\tablenotemark{a}} &\multicolumn{2}{c}{$\log{(L_{30\mu \textup{m--}1\mu \textup{m}})}$} &\multicolumn{2}{c}{$\log{(L_{1\mu \textup{m--}2\textup{keV}})}$} &\multicolumn{2}{c}{$\log{(L_{2\textup{keV--}10\textup{keV}})}$} &\multicolumn{2}{c}{$\log{(L_{10\textup{keV--}500\textup{keV}})}$}  }
\startdata
587745539970630036&  46.06&  0.03&  45.77&  0.07&  3.08&  0.25&  46.27&  0.03&  46.54&  0.01&  44.75&  0.17&  45.11&  0.11\\
587732772647993547&  45.35&  0.07&  44.98&  0.04&  2.59&  0.43&  45.52&  0.02&  45.76&  0.02&  44.24&  0.17&  44.6&  0.11\\
587735666930941958&  45.68&  0.08&  45.37&  0.06&  2.83&  0.51&  46.13&  0.02&  46.13&  0.02&  44.48&  0.17&  44.84&  0.11\\
587724198277808231&  46.31&  0.06&  46.39&  0.08&  2.31&  0.33&  46.78&  0.02&  46.67&  0.01&  44.97&  0.17&  45.33&  0.11\\
587729159520452814&  45.85&  0.05&  45.62&  0.06&  3.14&  0.36&  46.22&  0.02&  46.35&  0.02&  44.62&  0.17&  44.98&  0.11\\
588848900997054741&  44.71&  0.06&  44.53&  0.07&  3.62&  0.51&  45.02&  0.02&  45.27&  0.02&  43.79&  0.17&  44.15&  0.11\\
588017978340999239&  46.91&  0.06&  46.63&  0.03&  2.49&  0.33&  47.25&  0.02&  47.31&  0.01&  45.38&  0.18&  45.74&  0.12\\
587722984428798040&  46.29&  0.05&  46.07&  0.07&  3.33&  0.37&  46.47&  0.02&  46.81&  0.01&  44.92&  0.17&  45.28&  0.11\\
587732483289055296&  45.5&  0.06&  45.37&  0.04&  2.2&  0.3&  45.94&  0.01&  45.84&  0.01&  44.31&  0.17&  44.67&  0.11\\
587733081880789140&  45.54&  0.05&  45.3&  0.04&  2.64&  0.34&  45.91&  0.02&  45.96&  0.02&  44.39&  0.17&  44.76&  0.11
\enddata
\tablenotetext{a}{Taken over the range of 1\,$\mu$m to 2\,keV.}

\tablecomments{All luminosities are reported in $\log{(\mbox{erg s}^{-1})}$. This table is published in its entirety in the electronic edition of the online journal. A portion is shown here for guidance regarding its form and content.}
\end{deluxetable*}

\subsection{Normalization Wavelength}

Typically, bolometric corrections are computed with respect to $5100\,{\rm \AA}$ and we will report those values herein for backward compatibility with previous work.  However, we will generally report BCs relative to $2500\,{\rm \AA}$.  There are a number of reasons for this choice.  While $5100\,{\rm \AA}$ makes sense for low-redshift sources, especially when using the H$\beta$ line to estimate black hole masses and then the Eddington ratio from the ratio of $L_{\rm bol}$ to $M_{\rm BH}$, that rest-frame wavelength is inaccessible for the vast majority of SDSS quasars.  The SDSS quasar sample peaks at $z\sim1.5$, which corresponds to rest-frame spectral coverage of $\sim1500$--3700\,\AA.  As such, the majority of SDSS quasars have observed flux density measurements at $2500\,{\rm \AA}$.  For this reason, \citet{Richards06qlf} chose to $K$-correct to $z=2$ where the SDSS $i$-band roughly corresponds to $\lambda_{\rm eff}=2500\,{\rm \AA}$.  Moreover, $2500\,{\rm \AA}$ is typically used as the optical anchor point in the \luvaox\ relationship.  In addition to this, the $5100\,{\rm \AA}$ luminosity always has a larger host galaxy contamination.  As such, we have chosen to use $2500\,{\rm \AA}$ as our fiducial wavelength.

\subsection{Integrated Luminosities and Bolometric Corrections} \label{BCs}

For comparison with previous work, we compute the bolometric luminosity in a number of ways, including the tabulation of integrated optical and IR luminosities.    In terms of bolometric corrections, we start by computing BCs for each individual object in our sample, and, with those, compute mean BCs for the full sample.  In computing BCs we used SEDs constructed as discussed above: namely, using broadband observations where available, the \luvaox\ relationship, and ``gap-filling'' (with the mid-luminosity mean SED) as needed.  The overall mean \bctwofive\ and corresponding standard deviation values are $2.75 \pm 0.40$ using limits of \onemum\ and \twokev\ (hereafter called $L_{\rm disk}$, which avoids the issue of IR and hard X-ray double counting); at $5100\,{\rm \AA}$ this corresponds to $4.33 \pm 1.29$.  For comparison with \citet{Elvis94} and \citet{Richards06} we also determine the bolometric correction to $5100\,{\rm \AA}$ in the range of 30\,$\mu$m up to 10\,keV. We find \bcfiveone\ $=7.79 \pm 1.69$, as compared to the results of \citet{Elvis94} and \citet{Richards06} who found $11.8^{+12.9}_{-6.3}$ and $10.3 \pm 2.1$ respectively.  Using the range \onemum\ to 500\,keV from \citet{Marconi04} we find that \bctwofive\ $=2.97 \pm 0.43$.

Recent studies by \citet{Nemmen10} and \citet{Runnoe12} suggest the relationship between a quasar's monochromatic and bolometric luminosity is nonlinear.  As is expected given the nonlinear relationship used to connect the UV and the X-ray, we also find this to be the case; our best fit is a power-law of the form:
\begin{eqnarray}
 \log{(L_{{\rm Bol}})}=&&(0.9869 \pm 0.0003)\log{(\nu_{2500{\rm \AA}} L_{2500{\rm \AA}})}\nonumber \\
  	&&+ (1.051 \pm 0.014)
 \label{Lbol_v_Luv}
\end{eqnarray}
Note that while we allow for a nonlinear slope, the best-fit slope is very close to linear. The errors quoted on this fit are small because of our large sample size and only statistical, not systematic, errors are included.

\begin{figure}
\centering
 \includegraphics[width=3.5in]{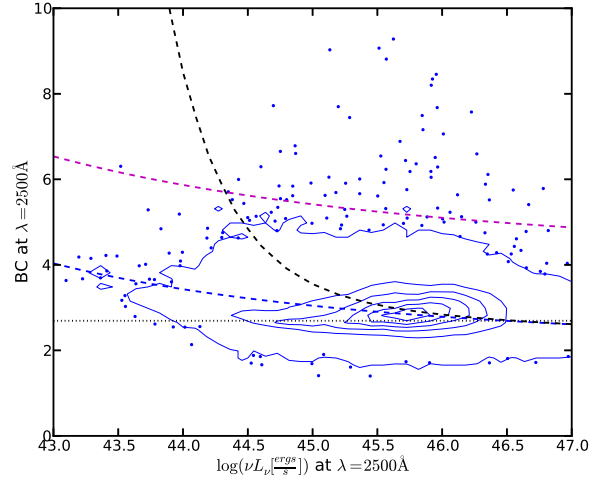}
 \caption{Disk bolometric correction as a function of UV luminosity, \ltwofive.  
The blue contours and scatter points shows a density plot for our quasar sample.  The contours indicate the linear density of scatter points on the plot.
The black dotted line is the bolometric correction for the overall mean SED (see Figure~\ref{mean_gap_cor}) without taking into account the nonlinear correlation between $L_{\rm UV}$ and $L_{\rm X}$. The blue dashed line shows the $L$-dependence of the BCs assuming the $L_{{\rm uv}}-L_{{\rm x}}$ relation \citep{Steffen06}; this line tracks the data points since the individual BCs are also dependent on $L_{\rm UV}$ .  The black dashed line shows the same relation with an extra bump added in the EUV (see Section~\ref{EUV}).   For comparison the relationship from \citet{Marconi04} is shown as the dotted maroon line scaled to the limits of integration and to 2500\AA.}
 \label{bc_v_lum}
\end{figure}

In addition to mean BCs, we give BCs and error estimates for each of the objects in our sample; these are shown by the points and contours in Figure~\ref{bc_v_lum}.  As did \citet{Marconi04}, we investigate how the BCs are dependent on quasar luminosity and, in turn, on the \luvaox\ relation that was assumed.  The black dotted line in Figure~\ref{bc_v_lum} shows the luminosity-independent BC that one gets for our mean SED and assuming a linear dependence between \luv\ and \lx (i.e. re-normalizing the mean SED without changing \aox).  To illustrate the importance of the observed nonlinear correlation between \luv\ and \lx\ we do the following.  We re-normalize our mean SED to \ltwofive\ values that span the range shown in Figure~\ref{bc_v_lum}, stepping in small increments.  We then truncate the mean SED at 1216\,\AA\ and connect it to \twohundredev\ using the \luvaox\ relationship and assuming $\alpha_{\rm x}=-1$ beyond \twohundredev.  We then calculate the resulting BCs over a range of luminosities; the results are given by the dashed blue line in Figure~\ref{bc_v_lum}.  
This line tracks the outliers well, but the core of our sample shows a weaker luminosity dependence to the BCs (as expected from Equation~(\ref{Lbol_v_Luv})).
This weaker luminosity dependence may reflect the trend of redder optical continua in low-luminosity quasars are bluer optical continua in high-luminosity quasars, which would counter-act the trends in \aox\ with \luv.

We have also included the line found by \citet[][maroon]{Marconi04} for comparison.  The \citet{Marconi04} line has been adjusted to match our limits of integration and the BC normalization wavelength.  We see that the form of the Marconi line and ours are similar,
and the offset is due to the fact that the Marconi SEDs use a shallower \luvaox\ relation that is extrapolated to \twokev\ (which would predict a soft X-ray excess in all quasars), whereas we use a steeper \luvaox\ relation extrapolated to \twohundredev.

The general shape of the distribution of individual BCs shows that the main luminosity dependence of our BCs is indeed due to the \luvaox\ relation.  We do, however, see a large spread of $\sim$1 on either side of the dashed line.
In particular, it is possible for a high-$L$ quasar to have a BC that is higher than a low-$L$ quasar even though the general trend is in the other direction.

In order to facilitate the determination of bolometric corrections using different limits of integration and normalizing wavelengths, in Table \ref{BC_table} we have tabulated the integrated luminosity and errors for individual SEDs over \thirtymum--\onemum, \onemum--\twokev, \twokev --\tenkev, and \tenkev--500\,keV.  The second range is our recommended range as it corresponds to $L_{\rm disk}$, but the sum of the first three ranges matches that used by \citet{Richards06} and the sum of the last three ranges is that used by \citet{Marconi04}. We further give \lfiveone, \ltwofive, and the resulting \bctwofive\ (relative to $L_{\rm disk}$).  We have not corrected for non-isotropic emission (i.e. the emitted light is not the same in all directions) in our tabulations; however, taking anisotropy into account, \citet{Runnoe12} suggest scaling the bolometric luminosities by 0.75 when calculating the bolometric luminosity over the range of \onemum--\tenkev.

\begin{figure}
\centering
 \includegraphics[width=3.5in]{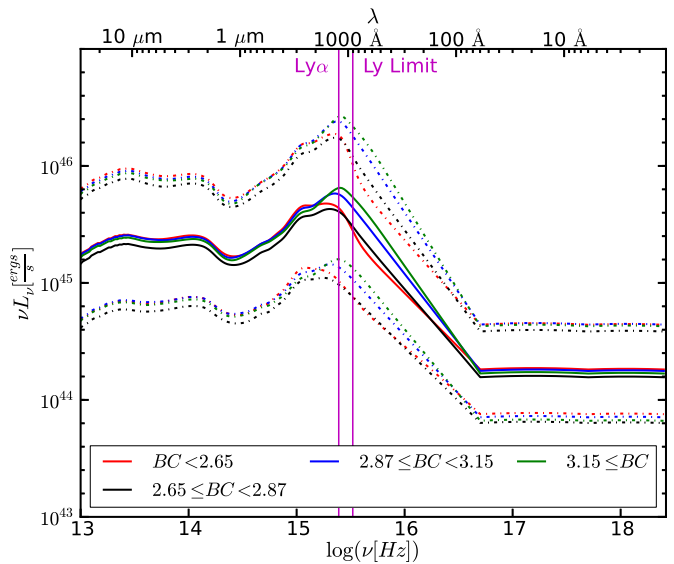}
 \caption{Mean SEDs for four bolometric correction bins. The higher BC SEDs show more emission in the UV--EUV part of the SED ($\lambda \lesssim 2000\,{\rm \AA}$) and an overall higher luminosity.  The steep drop-off of the low-BC SED shortward Ly$\alpha$ could be caused by unaccounted for dust reddening (see end of Section~\ref{BCs} for details). Figure~\ref{norm_seds} shows an expanded version of the UV section of this plot.}
 \label{bc_bin}
\end{figure}

With individual BCs for each quasar, we are able to construct BC-dependent mean SEDs using BC$_{\rm disk}$.  For this we split our sample into four equally populated bins: BC$<2.65$, $2.65\leq$BC$<2.87$, $2.87\leq$BC$<3.15$, and BC$\geq 3.15$, each containing $\sim29,000$ quasars. Unlike the luminosity- and \civ-dependent SEDs, these mean SEDs do not truncate the individual SEDs and connect the mean to the X-ray, but instead connects each individual SED to the X-ray and then takes the mean.  This is done since each SED needs to be connected to the X-ray in order to calculate the BC.  Figure~\ref{bc_bin} shows the resulting means and Figure~\ref{norm_seds}(d) shows the UV region when the SEDs are normalized at 2500\,\AA.  From these figures we see that the mean SEDs only differ significantly in the UV ($\lambda \lesssim 2000\,{\rm \AA}$) and the largely unknown regions of the SED in FUV could have significant effects as compared to the differences seen over 1000--2000\AA\ where the SED is well-measured.  
Although they were not included in the mean SEDs, the 11,468 quasars that show signs of significant dust reddening ($\Delta(g-i)>0.3$) all fall in the two lowest BC bins; this is expected since quasars with heavy dust reddening will appear to have a smaller BC.
This cut only accounts for strong dust reddening; if there is only a small amount of dust reddening then a high-BC quasar could easily fall into the low-BC bin. We might expect the low-BC bin to be contaminated by quasars with mild dust reddening, which could be the cause of the steep drop-off in the low-BC SED just past Ly$\alpha$.

\section{Discussion: The Unseen EUV Continuum} \label{EUV}

One of the reasons that we explored the mean SED in subsamples as a function of various quasar parameters is that we expect that the diversity of quasar properties (especially in broad emission lines) is a direct consequence of the diversity in SEDs \citep[e.g.,][]{Richards11}.  In particular, quasars with significant UV luminosity relative to the X-ray are capable of driving strong winds through radiation pressure on resonance line transitions \citep{Murray95,Proga00}.  Given the nonlinear relationship between \luv\ and \lx\ \citep[e.g.,][]{Steffen06},
one might expect to see significant differences in the SEDs of quasars as a function of luminosity.  \citet{Richards11} have argued that a potential wind diagnostic may be the properties of the \civ\ emission line in the context of the \citet{Boroson92} eigenvector 1 parameter space which looked at the difference between quasars using principal component analysis.  This was the main motivation for exploring the mean SED as a function of these parameters in Section~\ref{sed}.  While we saw significant differences in the mean SEDs as a function of luminosity, we saw fewer differences between the \civ\ composites than might have been expected.

While it is possible that the SEDs of quasar extrema (in UV emission) truly are similar, it is difficult to understand how BELR properties could be so different, yet have similar SEDs.  For example, there is a large range in the strength of the \ion{He}{2} 1640 line (with an ionization potential of 54.4 eV) in Figures~11 and 12 of \citet{Richards11}.  As such, we are led to a similar conclusion as \citet{Netzer79}, \citet{Korista97}, \citet{Done12}, and \citet{Lawrence12}--- namely that the EUV SEDs may be quite different from the standard power-law parameterization between the optical and X-ray.  This variation could be in a number of forms, including a second bump in the EUV or simply that the SED seen by the BELR is different from what we view \citep[e.g.,][]{Korista97}.

Here we suggest that the solution may be more than just a different mean SED in the EUV than is normally assumed, but rather that the shape of the EUV SED must be quite different for quasars at opposite extrema in terms of their BELR properties; i.e., there is {\em no} universal quasar SED (even after accounting for the \luvaox\ relationship)!  While \citet{Grupe10} find that only a factor of a few difference can be hidden in the EUV, their sample is restricted to low-redshift AGNs, whereas our suggestion is that the differences might be significant when considering the extremes of the distribution as spanned by the full SDSS quasar sample.  
In terms of a model where the BELR emission comes from both disk and wind components \citep{Collin98,Leighly04}, quasars with strong winds would have a weaker EUV continuum than quasars with strong disk components.  
What \citet{Richards11} refer to as disk-dominated objects have emission line features that are consistent with a much harder EUV SED than the wind-dominated objects \citep{Kruczek11}.  If the SEDs over the {\em observable} range of the EM spectrum are similar, the {\em unseen} part of the EM spectrum may yield very different SEDs. This hypothesis is consistent with radiation line driving being sensitive to the UV to ionizing flux ratio \citep{Proga00}.

If this hypothesis is correct, it would have important consequences for the determination of bolometric corrections (and thus $L/L_{\rm Edd} \propto L_{\rm Bol}/M_{\rm BH}$).  A full investigation into the range of EUV continuum properties is beyond the scope of this paper; however, herein we have created some examples showing how different these values might be for different assumptions of the EUV SED at the extrema of BELR properties.  

\begin{figure}
 \centering
 \includegraphics[width=3.5in]{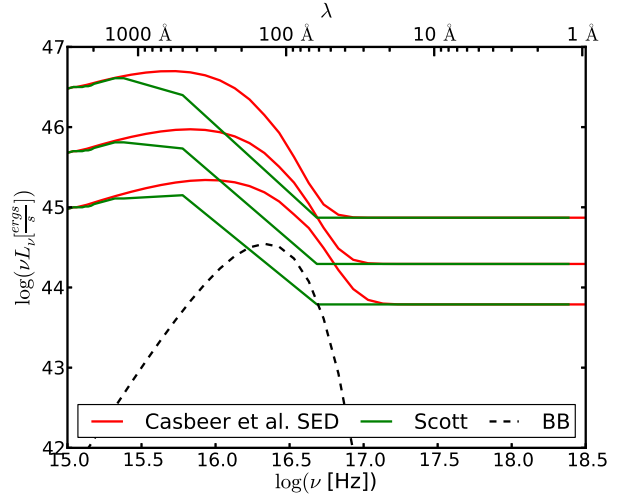}
 \caption{ The three models we used to represent the EUV part of our SEDs at three luminosities. The Casebeer et al. model is shown in red and the Scott$+$\luv--\lx\ model in green.  The black dashed line shows a blackbody with a peak energy of 23 eV and normalized so it peaks with $\log{(\nu L_{\nu})}=44.5$.
 These models show the differences between the observed and theoretical models shortward of Ly$\alpha$ as a function of luminosity and illustrate the relative level of our hypothetical extra EUV component.}
 \label{three_models}
\end{figure}

The most obvious deviation from our baseline prescription (using \luvaox\ to describe the unseen part of the SED) follows from the work of \citet{Scott04}
in the FUV.  The UV spectral index just shortward of Ly$\alpha$ was found to be different by \citet{Telfer02} and \citet{Scott04} ($\alpha_{{\rm FUV}}=-1.42$ as compared to \aox$=-1.54$ for the average quasar luminosity).  These differences can be explained by a luminosity-dependent spectral index given in Equation~(\ref{scott_eq}) \citep{Scott04} that is similar to that seen for $\alpha_{\rm ox}$ and by the fact that these two samples probe different ranges of \luv.   Thus, we create $L$-dependent model SEDs that do not just connect 2500\,\AA\ directly to \hundredev, but instead that connect 2500\,\AA\ first to 500\,\AA\ following Equation~\ref{scott_eq} and then connect 500\,\AA\ to \twohundredev\ using the \luvaox\ relationship to set the X-ray continuum level.  Such an SED is illustrated (green lines) in Figure~\ref{three_models} for three different luminosities, where it can be seen that these SEDs 
have a small (luminosity-dependent) excess of EUV flux as compared to a power-law fit between the optical and the X-ray.

\begin{figure*}
 \centering
  \begin{tabular}{cc}
 \includegraphics[width=3.2in]{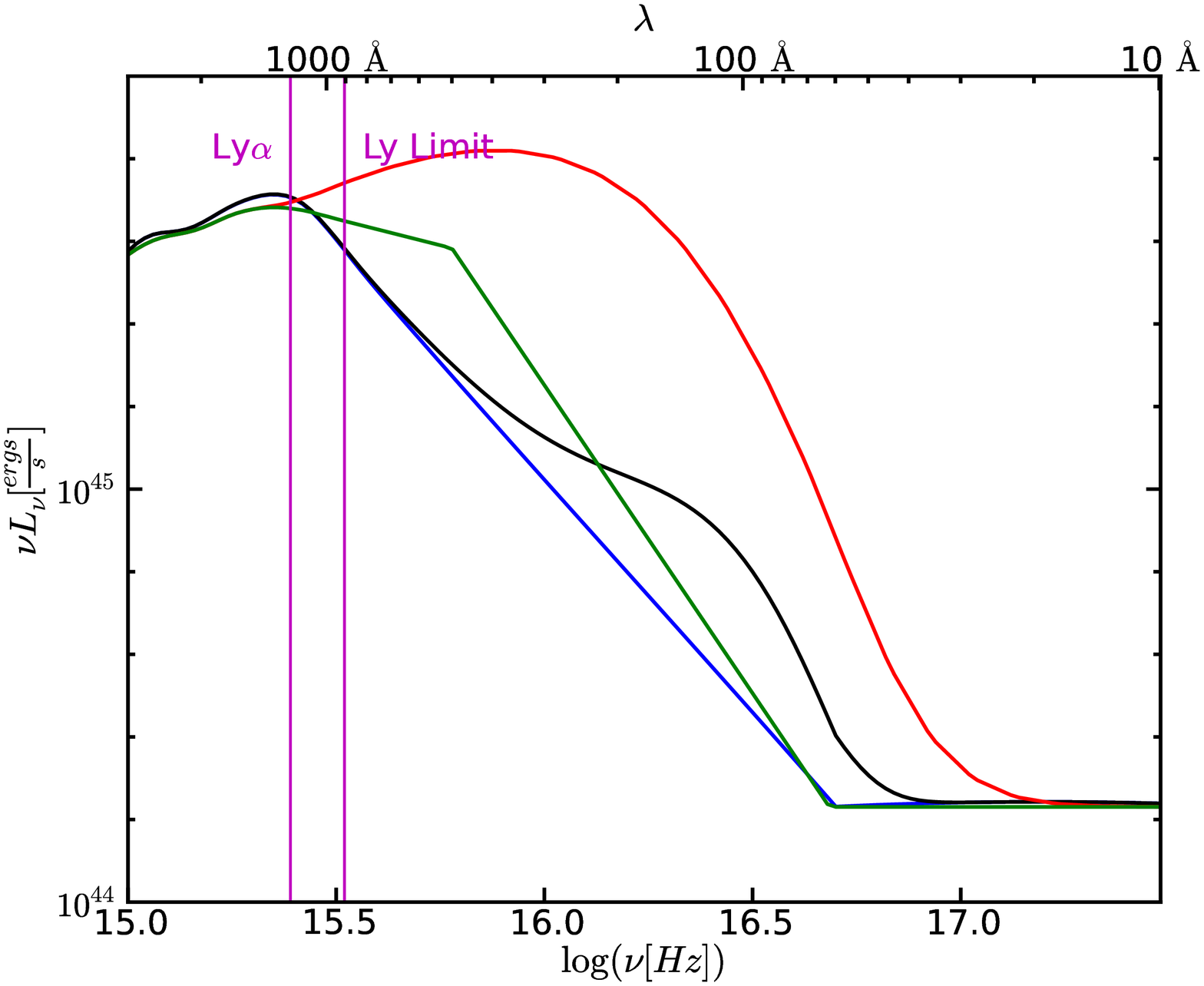} & \includegraphics[width=3.4in]{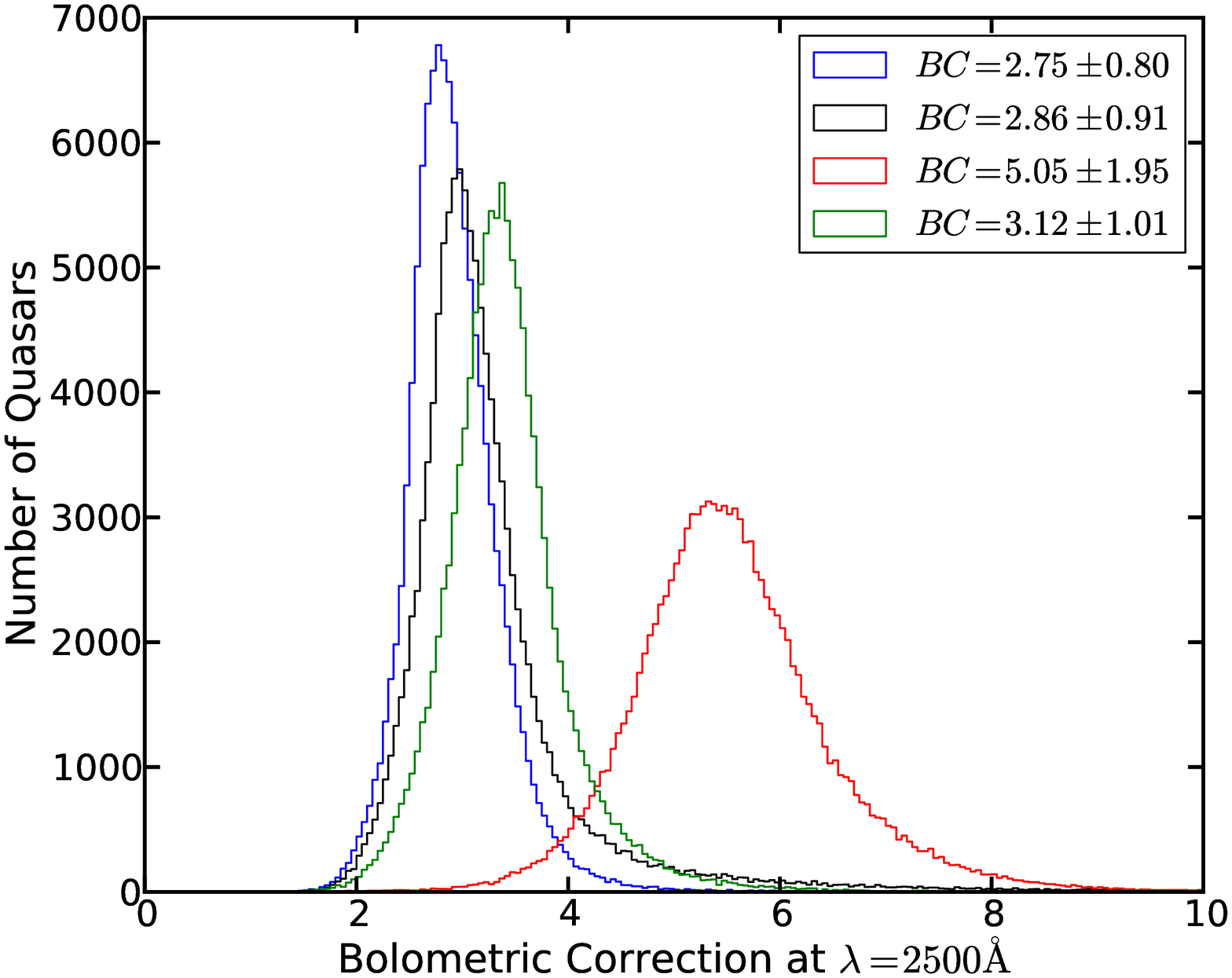} \\
 (a) EUV models & (b) Bolometric Corrections
 \end{tabular}
 
  \caption{(a) Mean SEDs for four different models: $L_{{\rm uv}}$--$L_{{\rm x}}$ from \citet[][blue]{Steffen06}, the $L_{{\rm uv}}$--$L_{{\rm x}}$ relation with a 23 eV blackbody with a peak value of $\log{(\nu L_{\nu})}=45$ (black), using the EUV extension from \citet[][green]{Scott04}, and the Casebeer et al. model \citep[][red]{Casebeer06}.
  (b) Bolometric corrections for the four models in (a) where the bolometric luminosity is taken from \onemum\ to \twokev\ ($L_{{\rm disk}}$). The errors in the legend give the 2$\sigma$ levels of the BC distributions.
  Models similar to all of those shown here are arguably well-justified and used in the literature. Their differences emphasize the need to better understand the distribution of EUV continua in quasars.}
  \label{bcs}
\end{figure*}

We compare this to model SEDs taken from \citet{Casebeer06} that 
are adapted from the CLOUDY \citep{Ferland02} so-called ``AGN" continuum.
The functional form of this continuum is given in \citet[][Equation (A1)]{Casebeer06}.  Essentially this SED consists of a power law
representing the optical/UV continuum with exponential cutoffs in the
infrared and UV, plus a power law in the X-rays, where the
normalization of the two components is set by $\alpha_{{\rm ox}}$ \citep[calibrated to ][]{Wilkes94}.  
\citet{Casebeer06} used a range of $\alpha_{{\rm ox}}$ and UV cutoffs to test the
influence of the SED on emission line ratios (and not to determine BCs); we
show in Figure \ref{three_models} the ones with $kT_{{\rm cut}}=50, 37, 27$ eV and $\alpha_{{\rm ox}}=-1.5, -1.57, -1.66$, from lowest to highest luminosity respectively (chosen to have the same optical luminosities as the Scott lines).

While we cannot directly measure the EUV part of the spectrum, it is interesting to compare the Scott and Casebeer et al. SEDs in the $\sim$800\,\AA\ range.  
For low-luminosity quasars, the data-driven Scott-based SED and the more theoretical (but empirically calibrated) Casebeer et al. SED (chosen to have the same \aox), follow the same upward trend.  This leads one to wonder if low-luminosity objects may indeed have much more EUV flux shortward of 500\AA\ (i.e., that the Casebeer et al. 2006 model is correct)
 than is typically assumed when applying the \luvaox\ prescription, as the EUV is unconstrained for the green curves between 500\,\AA\ and 0.2\,keV.  However, for the high-luminosity quasars, the Scott and Casebeer et al. SEDs do not follow the same trend even at $\sim800\,{\rm \AA}$, which suggests a model like the \citet{Casebeer06} SED likely overestimates the EUV emission in the most luminous quasars.
Indeed, a comparison of the UV emission lines in the extrema shown in Figure~12 of \citet{Richards11} suggests that the SEDs of quasars with ``disk''- and ``wind''-dominated BELRs may be very different in the range of $\sim$50\,eV where the ionization potentials of many UV lines lie \citep[e.g.,][Figure 13]{Richards11}.  

Since the wind-dominated objects are more luminous on average than the disk-dominated objects (as can be seen in Figure~\ref{civ_bin}(c)), we also consider a model where we add a blackbody component of fixed luminosity, such that it is significant in low-luminosity sources, but contributes little to the EUV continuum for high-luminosity sources.

To construct a toy model to illustrate such a situation, we start by extending our SEDs to the X-ray using the \citet{Steffen06} \luvaox\ relation and then added a blackbody peaking at 23 eV. 
This peak was chosen in order to add photons of the typical energy needed to produce the spectral lines we see (e.g., \civ\ and \ion{He}{2}).
 The normalization of the blackbody was taken to have $\log{(\nu L_{\nu})}=44.5$ at its peak (see the dashed black curve in Figure~\ref{three_models})  
so that the highest luminosity quasars remain unaffected but the lower luminosity quasars gain an extra feature which has the correct sign to explain the broad range of EWs for, e.g., \civ\ and \ion{He}{2}
\citep{Richards11}.  This model leads to a 
larger spread in their bolometric corrections than 
when no extra component is used (see Figure~\ref{bcs}(b)).  While we lack direct observational evidence for such a component,
the range of emission line strengths for lines with ionization potentials of $\sim$50\,eV, suggestion that some, but not all, quasars have a need for more ionizing flux in this region of the EM spectrum in order to understand their emission line properties.  We include it here as a way to illustrate how significantly such a component would change the bolometric corrections of low-luminosity quasars.  

We compare these four models, which were chosen to bracket the range of reasonable EUV shapes for a given $\alpha_{\rm ox}$, shown in Figure~\ref{bcs}(a) for a single luminosity.   Figure~\ref{bcs}(b)  then shows how the BC distribution would change under the assumption of each of these models.  The biggest deviation from the standard model results when using the Casebeer et al. SEDs. 
These SEDs may have too much EUV flux for high-luminosity sources, while being more reasonable at low-luminosities
(see Figure~\ref{three_models}).  On the other hand, the Scott-based and extra EUV component SEDs depict a more subtle shift in the mean BCs.  More importantly, however, is that these deviations from our standard SED produce BC changes that are {\em systematic} with \luv.    
Although there is a lack of data in the EUV, in Figure~\ref{bc_v_lum}, the dashed black line shows how much we would expect the BCs to change for low-luminosity sources if there were an extra EUV continuum component.  Adding the $L$-dependence from Scott to this model would make this contrast even stronger.  

While this extra EUV component is speculative, the ramifications are such that it is important to consider the possibility.  Moreover, the $L$-dependence of the FUV continuum based on the work of \cite{Scott04} already suggests that a correction in this direction is needed.  Achieving a better understanding of the EUV continuum is clearly of importance for understanding quasars physics, as a change of a factor of a few in BCs for low-luminosity sources translates to the same correction factor for both \lbol\ and the Eddington ratio (i.e., accretion rates).  Ideally, we would also like to consider the BC distribution as a function of quantities like the accretion rate, but such analysis is difficult without first understanding any systematic effects in the unseen EUV part of the spectrum.

\section{Conclusions} \label{conclusions}

We have compiled a sample of 119,652 quasars detected in the SDSS.  These data are supplemented with multi-wavelength data spanning from the mid-IR through the UV (Table~\ref{data_table}).  This data set was used to construct a new mean SED consisting of 108,184 non-reddened quasars, with rest frame coverage from $\sim 20\, \mu$m -- $912\,{\rm \AA}$ (Table~\ref{mean_sed_table}). 
By splitting our sample into luminosity bins we constructed three luminosity-dependent mean SEDs and found $\alpha_{{\rm FUV}}$ to be dependent on luminosity in the sense that more luminous quasars have redder $\alpha_{{\rm FUV}}$ continua (Figure~\ref{norm_seds}(b)), 
consistent with \citet{Scott04}. In addition, the high-luminosity quasars also show signs of having bluer optical continua (Figure~\ref{norm_seds}(b)) and more hot dust emission than the low-luminosity quasars (Figure~\ref{norm_seds}(a)).  
When splitting our sample based on \civ\ properties we saw differences in the Balmer continuum, Ly$\alpha$ and \civ\ (by construction) line strengths (Figure~\ref{norm_seds}(c)) that are consistent with eigenvector 1 trends \citep{Brotherton99}.

We also constructed SEDs for each quasar and, from those, found bolometric corrections (Table~\ref{BC_table}).  The overall mean is \bctwofive $=2.75 \pm 0.40$ using integration limits of \onemum-- \twokev. 
While the range of bolometric corrections indicated by the IR through NUV data alone is reasonably small, there can be significant changes in the distribution of bolometric corrections when different models are assumed in the unseen EUV part of the SED (Figure~\ref{bcs}(b)).  Although more work has to be done to determine which model should be used in the EUV, it is nevertheless clear that it is important to consider potentially significant differences in the EUV part of the SED at the extrema of quasar properties (e.g., in luminosity and in eigenvector 1 parameter space).
In future work, we will further explore the cause(s) of the width in the bolometric correction distribution (after correcting for known luminosity trends) and will characterize the systematic error in the accretion rate estimates.

\acknowledgments

G.T.R. acknowledges support from an Alfred P. Sloan and an Alexander von Humboldt Research Fellowship along with NASA grants NNX08AJ27G, NNX10AF74G, and NNX12AI49G.  
This work is based in part on observations made with the {\em Spitzer Space Telescope}, which is operated by the Jet Propulsion Laboratory, California Institute of Technology under a contract with NASA. Support for this work was provided by NASA through an award issued by JPL/Caltech.
We thank both David Schiminovich and David Hogg for providing {\em GALEX} forced photometry and both Anna Sajina and the referee for critical review of the manuscript.

\bibliography{ms}
\end{document}